\documentclass{article}
\usepackage[utf8]{inputenc}

\usepackage{amssymb,amsthm,amsmath,mathrsfs,mathtools}
\usepackage{booktabs}
\usepackage{multirow}
\usepackage{siunitx}
\usepackage{amsfonts}
\usepackage{enumitem}
\usepackage{graphicx,placeins}
\usepackage{lineno}
\usepackage{hyperref}
\hypersetup{
	colorlinks=true,
	linkcolor=blue,
	filecolor=blue,      
	urlcolor=blue,
}
\DeclareUnicodeCharacter{0306}{ }

\theoremstyle{plain}
\newtheorem{theorem}{Theorem}
\newtheorem{lemma}{Lemma}
\newtheorem{corollary}{Corollary}

\theoremstyle{definition}
\newtheorem{definition}{Definition}
\renewcommand{\labelenumi}{$\mathrm{(\roman{enumi})}$}

\usepackage{algpseudocode,algorithm,algorithmicx}

\algrenewcommand\algorithmicrequire{\textbf{Input:}}
\algrenewcommand\algorithmicensure{\textbf{Output:}}
\algdef{SE}[DOWHILE]{Do}{doWhile}{\algorithmicdo}[1]{\algorithmicwhile\ #1}%

\usepackage[backend=biber, isbn=true, giveninits=true,maxbibnames=99,date=year,url=false, sorting=none,style=nature]{biblatex}
\renewbibmacro*{doi+eprint+url}{%
	\printfield{doi}%
	\newunit\newblock%
	\iftoggle{bbx:eprint}{%
		\usebibmacro{eprint}%
	}{}%
	\newunit\newblock%
	\iffieldundef{doi}{%
		\iffieldundef{eprint}{
			\usebibmacro{url+urldate}}%
		{}%
	}{}}
\usepackage{geometry}
\title{Integral Equations \&  Model Reduction For Fast Computation of Nonlinear Periodic Response}
\author{Gergely Buza, George Haller, Shobhit Jain\footnote{Corresponding author (shjain@ethz.ch)}}
\date{\vspace{-5ex}
}

\begin{document}
	\maketitle
	\begin{center}
		Institute for Mechanical Systems, ETH Zürich \\
		Leonhardstrasse 21, 8092 Zürich, Switzerland 
		\par\end{center}
	\begin{abstract}
		We propose a reformulation for the integral equations approach of Jain, Breunung \& Haller [\emph{Nonlinear Dyn.} 97, 313--341 (2019)] to steady-state response computation for periodically forced nonlinear mechanical systems. This reformulation results in additional speed-up and better convergence. We show that the solutions of the reformulated equations are in one-to-one correspondence with those of the original integral equations and derive conditions under which a collocation type approximation converges to the exact solution in the reformulated setting. Furthermore, we observe that model reduction using a selected set of vibration modes of the linearized system substantially enhances the computational performance. Finally, we discuss an open-source implementation of this approach and demonstrate the gains in computational performance using three examples that also include nonlinear finite-element models.
	\end{abstract}

	\section{Introduction}
	Computing the steady-state response of  periodically forced nonlinear systems is a challenging task for contemporary engineering problems comprising high-dimensional finite element models. A number of methods are nominally available in the literature for nonlinear periodic response calculation, ranging from analytical perturbation techniques \cite{Nayfeh1974NonlinearAO,nayfeh1} to standalone computational packages (AUTO~\cite{Auto}, \textsc{coco}~\cite{Dankowicz2013a}, NLvib~\cite{Krack2019}) that perform numerical continuation (see \cite{Renson2016,Jain2019} for a review).
	Despite today's advances in computing, however, a good approximation to nonlinear \emph{forced response curves} in complex structural vibration problems remains challenging to obtain and hence model reduction is still required~\cite{Ponsioen2019ExactMR}.
	
	In this work, we focus on the recently proposed integral equations approach~\cite{Jain2019} to the computation of steady-state response in nonlinear mechanical systems. Showing superior computational performance over other methods, this approach uses an explicit Green's function in the second-order form to derive an integral equation, whose solution represents the steady-state response. This solution is then computed via collocation or spectral methods. A distinguishing aspect of this approach is that it allows for the simple and fast Picard iteration in obtaining the steady-state response even for non-smooth mechanical systems. From a computational perspective, this circumvents the computation and inversion of the Jacobian matrices which is computationally intensive for high-dimensional problems. At the same time, however, the Picard iteration may not converge near external resonances~\cite{Jain2019}. Near such resonances, one is therefore forced to switch to a more expensive Newton–Raphson scheme to secure convergence.

	The objectives of this paper are two-fold. First, we use a reformulation of the original integral equation approach, which leads to improved speed and convergence. This reformulation is motivated by an idea of Kumar and Sloan~\cite{kumar87} for scalar Hammerstein-type integral equations. Their approach reorders the nonlinearity and integration operations to obtain better computational performance using a collocation-type approximation.  Kumar and Sloan~\cite{kumar87} proved a one-to-one correspondence between the solutions of the original and reformulated integral equations. Furthermore, they specified the rate of convergence of the collocation approximation to the exact solution when using this new, reformulated approach.
	In this work, we extend these results to vector-valued functions before applying them to the integral equations presented in \cite{Jain2019}. We further verify the conditions under which Picard iteration is guaranteed to converge in the reformulated setting and implement both Picard and Newton--Raphson iterations in an open-source package~\cite{SST}.
	
	Second, we approach the reformulated integral equation method from a model reduction perspective, given that reduced-order models (ROM) are still required to cope with the complexity of contemporary engineering structures. Projection-based ROMs, for instance, are constructed by projecting the governing equations to a linear subspace which may be identified using a variety of techniques. Due to the general lack of invariance of such linear subspaces in nonlinear systems, there are no mathematical results confirming the relevance of projection-based ROMs for nonlinear model reduction. Yet such projection-based techniques are a common choice for model reduction due to their simple implementation. A recently developed technique~\cite{Buza2020} allows us to optimally identify modal subspaces for projection-based reduction using the rigorous theory of spectral submanifolds (SSM)~\cite{Haller2016}. SSMs are the smoothest nonlinear continuations of linear modal subspaces that are invariant under the nonlinear flow and allow the reduction of the nonlinear dynamics into an exact, lower-dimensional invariant manifold in the phase space. In this work, we also equip the reformulated integral equations approach with the SSM-based model reduction procedure~\cite{Buza2020}, which allows us to compute the steady-state periodic response of finite-element problems in a fast, automated and reliable manner.

	The remainder of this paper is organized as follows. We discuss the reformulation of the integral equation in Section~\ref{sect:reform} after a short introduction to the general setup in Section \ref{sect:setup}. Section~\ref{sectionapprox} deals with numerical analysis of the proposed approach. Here, we discuss the numerical advantages of the reformulated integral equation relative to the original one, followed by results on numerical convergence and iterative solution methods, i.e., Picard iteration and Newton--Raphson iteration.  Finally, in Section \ref{chapter4}, we use an open-source implementation~\cite{SST} of the results  to illustrate the improvements arising from the proposed reformulation equipped with model reduction on three mechanical examples.

	\section{Setup}
	\label{sect:setup}
	
	We consider mechanical systems with geometric nonlinearities of the form
	\begin{equation}
		M \ddot{x} + C \dot{x} + K x + S(x) = f,
		\label{dynsys}
	\end{equation}
	where $x(t) = (x_1(t), \ldots, x_n(t)) \in  \mathbb{R}^n$ is the vector of generalized coordinates; $M \in \mathbb{R}^{n\times n}$ is the positive definite mass matrix;
	$C \in \mathbb{R}^{n\times n}$ is the damping matrix; $K \in \mathbb{R}^{n\times n}$ is the positive semi-definite stiffness matrix;
	$S$ is a nonlinear, Lipschitz continuous function satisfying $S(x) = \mathcal{O} \left( \vert x \vert^2  \right)$ with Lipschitz constant $L_S$, where we denote by $\vert \cdot \vert$ the standard Euclidean norm;  and $f = (f_1,\ldots , f_n)$ is a time-dependent, $T$-periodic forcing.
	
	We assume proportional damping, i.e., that the damping matrix $C$ can be expressed as a linear combination of the matrices $M$ and $K$. This allows us to decouple the full system~\eqref{dynsys} at the linear level using the undamped vibration modes $u_j \in \mathbb{R}^n $, defined by the eigenvalue problem
	\begin{equation}
		\left( K - \omega^2_{0,j} M \right) u_j = 0, \qquad \text{for} \quad j = 1, \dots, n,
		\label{frequencyeq}
	\end{equation}
	where $\omega_{0,j}$ is the eigenfrequency of $u_j$.
	Using the linear transformation $x = U \eta$, where $\eta \in \mathbb{R}^n $ denotes the modal variables, and $U = [u_1, \dots , u_n] \in \mathbb{R}^{n\times n}$  is the transformation matrix composed of the undamped vibration modes,  we transform the  original system~\eqref{dynsys} as
	\begin{equation}
		U^{\top} M U \ddot{\eta}(t)+U^{\top} C U \dot{\eta}(t)+ U^{\top} K U \eta(t) + U^{\top} S(U \eta(t)) = U^{\top} f(t),
		\label{modaldecomp}
	\end{equation}
	where $(\bullet)^{\top}$ denotes the matrix transpose. Note that the $n$ equations in system~\eqref{modaldecomp} decouple at the linear level but become generally coupled under the nonlinear term $U^{\top} S(U \eta(t))$.
	
	\subsection{Modal truncation}
	\label{modaltruncation}
	The general idea of \textit{modal truncation} is to project the equations of motion~\eqref{dynsys} onto linear subspaces spanned by the vibration modes~\cite{geradin2015mechanical}. Given a truncated set of $m$ modes, say, $ u_1, \dots, u_m$  ($m \leq n$), we obtain a ROM from the full system~\eqref{dynsys} using the reduced transformation matrix $U_m = [u_1, \dots , u_m] \in \mathbb{R}^{n \times m}$ via Galerkin projection as
	\begin{equation}
		U_m^{\top} M U_m \ddot{\eta}(t)+U_m^{\top} C U_m \dot{\eta}(t)+ U_m^{\top} K U_m \eta(t) + U_m^{\top} S(U_m \eta(t)) = U_m^{\top} f(t),
		\label{modelreduct}    
	\end{equation}
	where the ROM~\eqref{modelreduct} has only $m$ unknowns ($\eta \in \mathbb{R}^{m}$). Note that we recover the full system~\eqref{modaldecomp} when all modes are included in $U_m$, i.e., when $m=n$. 
	
	An optimal set of modes for the above truncation can be chosen using the mode-selection criterion in~\cite{Buza2020}, as mentioned in the Introduction. This criterion is based on the recent theory of Spectral Submanifolds (SSMs)~\cite{Haller2016}, which are exact invariant manifolds that act as nonlinear continuations of linear normal modes in the phase space. Starting with an initial set of modes obtained from linear mode superposition, this criterion systematically identifies the modes whose associated SSMs have the largest local curvature. Such modes are the most senstitive to system nonlinearities and hence most relevant for any projection-based ROM. The mode selection process is automated for general systems (see Section 5 in \cite{Buza2020}) and has a open-source implementation~\cite{SST}, which we employ for modal truncation in this work.
	
	Next, we discuss how the steady-state response to periodic forcing can be obtained from such a truncated set of modes using the integral equations approach of Jain, Breunung \& Haller~\cite{Jain2019}. 
	
	\subsection{Integral equations for steady-state response}
	Without loss of generality, we assume that the undamped vibration modes are mass-normalized, i.e., $U^TMU=I$ and in particular, $U_m^T M U_m = I$. For notational purposes, we write the linear part of the $j-$th equation  in system~\eqref{modelreduct} as
	\begin{equation}
		\ddot{\eta}_j(t) + 2 \zeta_j \omega_{0,j} \dot{\eta}_j(t) + \omega_{0,j}^2 \eta_j(t)=\varphi_j(t), \qquad j = 1,\dots,m,
		\label{linearsys}
	\end{equation}
	where
	$
	\omega_{0,j}=\sqrt{\langle u_j,Ku_j \rangle}
	$
	are the undamped natural frequencies,
	$
	\zeta_j = \langle u_j,Cu_j \rangle / (2 \omega_{0,j})
	$
	are the modal damping coefficients,
	and
	$
	\varphi_j(t) = \langle u_j,f(t) \rangle
	$
	are the modal participation factors.
	
	We arrange the eigenvalues of the damped linear system~\eqref{linearsys} as 
	\begin{equation}
		\lambda_{2j-1,2j}=\left( -\zeta_j \pm \sqrt{\zeta_j^2-1} \right) \omega_{0,j}, \qquad \text{for} \quad
		j = 1,\dots,m.
		\label{eigenvalues}
	\end{equation}
	As in \cite{Jain2019}, we further define the constants
	\begin{equation}
		\alpha_j := \mathrm{Re}(\lambda_{2j}), \qquad \omega_j : = \vert \mathrm{Im}(\lambda_{2j}) \vert,
		\qquad \beta_j := \lambda_{2j-1}, \qquad \gamma_j = \lambda_{2j}, \qquad j = 1,\dots,m.
	\end{equation}
	For a $T$-periodic forcing $f$, the following statement recalls the second-order Green's function (see Lemma~3 in \cite{Jain2019}) to compute the periodic response of the reduced linear system~\eqref{linearsys}.
	
	\begin{lemma}
		For a $T$-periodic forcing $f$, if the non-resonance conditions
		\begin{equation}
			\lambda_j \neq i \frac{2 \pi}{T}l, \qquad l \in \mathbb{Z}
		\end{equation}
		are satisfied for all eigenvalues $\lambda_1,\dots,\lambda_{2m}$ defined in \eqref{eigenvalues},
		then there exists a unique $T$-periodic response for the linear system~\eqref{linearsys}, given by
		\begin{equation}
			\eta_{lin} (t) = \int_0^T L_m(t-s,T)U_m^{\top} f(s) \mathrm{d}s,    
			\label{linearresponse}
		\end{equation}
		where $L_m(t,T)$ is the diagonal Green's function matrix for the modal displacement variables defined as
		$$
		L_m(t,T) = \mathrm{diag} \big( 
		L_1(t,T),\dots,L_m(t,T)
		\big) \in \mathbb{R}^{m\times m},
		$$
		with 
		\begin{equation}
			L_j(t,T) =
			\begin{cases}
				\frac{e^{\alpha_j t}}{\omega_j} \left[ \frac{e^{\alpha_j T} \left[ \sin \omega_j(t+T)-e^{\alpha_j T} \sin \omega_j t \right] }{1+e^{2 \alpha_jT}-2 e^{\alpha_jT} \cos \omega_j T} +h(t) \sin \omega_j t \right], & \mathrm{if } \quad \zeta_j<1,\\
				\frac{e^{\alpha_j(t+T)} \left[ \left(1-e^{\alpha_j T} \right)t+T \right]}{\left(1-e^{\alpha_j T} \right)^2}+h(t)t e^{\alpha_jt}, & \mathrm{if } \quad \zeta_j=1,\\
				\frac{1}{(\beta_j - \gamma_j )} \left[
				\frac{e^{\beta_j(t+T)}}{1-e^{\beta_jT}}-
				\frac{e^{\gamma_j(t+T)}}{1-e^{\gamma_jT}}+
				h(t) \left( e^{\beta_j t}-e^{\gamma_j t} \right)
				\right], & \mathrm{if } \quad \zeta_j>1,
			\end{cases}
			\qquad j = 1, \dots , m,
			\label{greenfunct}
		\end{equation}
		where $h(t)$ denotes the Heaviside step function.
	\end{lemma}
	\begin{proof}
		The proof carries over directly from the proof of Lemma 3 in \cite{Jain2019} if we simply replace $n$ with $m$ since the system~\eqref{linearsys} consists of decoupled equations.
	\end{proof}
	
	The Green's function~\eqref{greenfunct} of the linear system~\eqref{linearsys} provides us with an integral equation, whose solution represents the nonlinear periodic response of the reduced nonlinear system~\eqref{modelreduct}, as follows.
	
	\begin{theorem}
		\begin{enumerate}
			\item If $\eta(t) $ is a $T$-periodic solution of the nonlinear system \eqref{modelreduct}, then $\eta(t)$ must satisfy the integral equation
			\begin{equation}
				\eta(t) = \eta_{lin} (t) -  \int_0^T L_m(t-s,T) U_m^{\top} S \left( U_m \eta (s) \right) \mathrm{d}s,
				\label{inteq}
			\end{equation}
			with $\eta_{lin}$ and $L_m$ defined in \eqref{linearresponse} and \eqref{greenfunct}.
			\item Furthermore, any continuous, $T$-periodic solution $\eta(t)$ of \eqref{inteq} is a $T$-periodic solution of the nonlinear system \eqref{modelreduct}.
		\end{enumerate}
		\label{inteqthm}
	\end{theorem}
	\begin{proof}
		This is a special case of Theorem 3 in \cite{Jain2019}, when applied to system~\eqref{modelreduct}.
	\end{proof}
	
	Once again, note that the integral equation~\eqref{inteq} provides us the nonlinear periodic response of the full system~\eqref{modaldecomp} when $m=n$. As discussed in \cite{Jain2019}, the integral equation formulation has advantages in the computation of the nonlinear steady-state response of mechanical systems. We aim to further reduce computational costs of this integral equation approach by using a reformulation due to Kumar \& Sloan ~\cite{kumar87}, which we discuss in the next section. 
	
	\section{Reformulation of the integral equation}
	\label{sect:reform}
	Kumar and Sloan~\cite{kumar87} established a one-to-one correspondence between the solutions $y(t)$ of scalar Hammerstein-type integral equation of the form 
	\begin{equation}
		y(t) = f(t) + \int_a^b K(t,s) g \left( s,y \left(s\right) \right) \mathrm{d}s, \qquad t \in \left[ a, b \right],
	\end{equation}
	with the solutions $z(t)$ of the integral equation
	\begin{equation}
		z(t) = g \left( t, f(t) + \int_a^b K(t,s) z(s) \, \mathrm{d}s  \right), \qquad t \in \left[ a, b \right],
	\end{equation}
	where $-\infty <a<b< \infty$, $f:\left[ a, b \right] 	\rightarrow \mathbb{R} $,
	$K: \left[ a, b \right] \times \left[ a, b \right] \rightarrow \mathbb{R} $,
	and $g: \left[ a, b \right] \times \mathbb{R} \rightarrow \mathbb{R}$ are known scalar functions (see Lemma 1 in~\cite{kumar87}). In Appendix~\ref{appendix2}, we extend their results to vector-valued functions, i.e., $f:\left[ a, b \right] 	\rightarrow \mathbb{R}^n $,
	$K: \left[ a, b \right] \times \left[ a, b \right] \rightarrow \mathbb{R}^{n\times n} $,
	and $g: \left[ a, b \right] \times \mathbb{R}^n \rightarrow \mathbb{R}^n$.  This allows us to reformulate the integral equation~\eqref{inteq} in the following equivalent form.
	
	\begin{theorem}
		Any solution $\eta(t)$ of the integral equation~\eqref{inteq} is in one-to-one correspondence with a solution $\zeta(t)$ of the integral equation
		\begin{equation}
			\zeta(t) = -U_m^{\top} S \left(U_m \left[\eta_{lin}(t) + \int_0^T L_m(t-s,T)  U_m^{\top}\zeta (s) \right] \mathrm{d}s\right),
			\label{reforminteq}
		\end{equation}
		such that 
		\begin{equation}
			\eta(t) = \eta_{lin}(t) + \int_0^T L_m(t-s,T)  \zeta (s) \mathrm{d}s,
		\end{equation}
		and 
		\begin{equation}
			\zeta(t) = -U_m^{\top} S \left(U_m \eta(t) \right).
		\end{equation}
		\label{refromthm}
	\end{theorem}
	\begin{proof}
		The proof is an application of Lemma~\ref{kumarsloan} in Appendix~\ref{appendix2} (which extends Lemma~1 in \cite{kumar87} for vector-valued functions) to integral equations~\eqref{inteq} and ~\eqref{reforminteq}.
	\end{proof}

	As discussed by Kumar \& Sloan~\cite{kumar87}, one advantage of the reformulation in Theorem~\ref{refromthm} is that the convolution integral in eq.~\eqref{reforminteq} becomes independent of the exact shape of $\zeta$ when a collocation type approximation is implemented. Thus, when solving eq.~\eqref{reforminteq} via iterative schemes, one only needs to compute the convolution integral once in contrast to \eqref{inteq}, for which the integral has to be evaluated at each iteration step. We discuss this and other advantages of the proposed reformulation in the following sections.
	
	\section{Numerical analysis}
	\label{sectionapprox}
	In order to find the steady-state response of the dynamical system \eqref{dynsys}, we solve the integral equation~\eqref{inteq}~\cite{Jain2019} or, equivalently, the reformulated integral equation~\eqref{reforminteq} via a numerical approximation. We first compare the numerical approximation to the solution of equations \eqref{inteq} and \eqref{reforminteq}. 
	\subsection{Numerical comparison}
	\label{sectioncomparison}
	We use a collocation-type approximation to the solution $\eta(t)$ of eq.~\eqref{inteq} in the form
	\begin{equation}
		\eta^N(t) = \sum_{i=1}^N a_i v^N_i (t), \qquad t \in [0,T],
		\label{approx}
	\end{equation}
	with $N$ collocation points in the interval $[0,T]$, known basis functions $v^N_i$, and unknown coefficients $a_i \in \mathbb{R}^m $. Substituting eq.~\eqref{approx} into eq.~\eqref{inteq}, and evaluating it at each of the collocation points $t_1, \dots , t_N$, we obtain a closed system of nonlinear equations in terms of the coefficients $a_i$ as 
	\begin{equation}
		\sum_{i=0}^N a_i v^N_i (t_j) = \eta_{lin}(t_j) - \int_0^T L_m(t_j-s,T) U_m^{\top} S \left( U_m \sum_{i=1}^N a_i v^N_i (s)  \right) \mathrm{d}s, \qquad j = 1, \dots,N.
		\label{yapprox}
	\end{equation}
	
	Consider now the collocation approximation to $\zeta(t)$ in eq.~\eqref{reforminteq} as
	\begin{equation}
		\zeta^N(t) = \sum_{i=1}^N b_i v^N_i (t), \qquad t \in [a,b].
		\label{approxz}
	\end{equation}
	As in the previous case, we substitute eq.~\eqref{approxz} into eq.~\eqref{reforminteq} and evaluate it at every collocation point $t_1, \dots , t_N$ to determine the unknown coefficients $b_i \in \mathbb{R}^m $ as
	\begin{equation}
		\sum_{i=1}^N b_i v^N_i (t_j) = -U_m^{\top} S \left(U_m \eta_{lin}(t_j)  + \sum_{i=1}^N U_m\left[\int_0^T L_m(t_j-s,T)  U_m^{\top} v^N_i (s) \right] \mathrm{d}s~b_i\right), \qquad j = 1, \dots,N.
		\label{zzapprox}
	\end{equation}
	In general, eq.~\eqref{yapprox} or eq.~\eqref{zzapprox}  require an iterative solution method (e.g., Picard iteration or Newton-Raphson iteration) due to the presence of the nonlinearity $S$. As noted by Kumar \& Sloan~\cite{kumar87}, a disadvantage of applying the collocation approximation to the original integral equation~\eqref{inteq} is that the $N$ integrals in \eqref{yapprox} have to be evaluated at every step of the iteration. In contrast, we observe from the approximation~\eqref{zzapprox} of the reformulated integral eq.~\eqref{reforminteq} that the convolution integrals become independent of the current iteration step (i.e., of the coefficients $b_i$) and thus they only need to be computed once.
	
	\subsection{Numerical convergence}
	In the following, we show that $\zeta^N (t)$ converges to the exact solution $\zeta(t)$ of eq.~\eqref{reforminteq}, and analyze the rate of this convergence analogous to Section 4 in \cite{kumar87}.
	
	Let the basis functions $v^N_i$ be piecewise polynomial on $[0,T]$. These basis functions define a finite-dimensional subspace of $C \left( [0,T], \mathbb{R}^m  \right)$ as 
	\begin{equation}
		V^N:= \mathrm{span} \left\{ v^N_1, \dots , v^N_N \right\} \otimes \mathbb{R}^m \subset C \left( [0,T], \mathbb{R}^m  \right).
		\label{V_N}
	\end{equation}
	Let $P^N: C \left( [0,T], \mathbb{R}^m  \right) \rightarrow V^N$ be the interpolatory projection operator defined by
	\begin{equation}
		P^N w:= \sum_{i=1}^N w(t_i) v^N_i, \qquad w \in C \left( [0,T], \mathbb{R}^m  \right),
		\label{interpop}
	\end{equation}
	which assigns to any continuous function $w$ its piecewise polynomial interpolant. Furthermore, we define a substitution operator $\mathcal{G}:C \left( [0,T], \mathbb{R}^m  \right) \rightarrow C \left( [0,T], \mathbb{R}^m  \right)$ and an affine integral operator $\mathcal{T}:C \left( [0,T], \mathbb{R}^m  \right) \rightarrow C \left( [0,T], \mathbb{R}^m  \right)$ as
	\begin{align}
		\label{G}
		\mathcal{G}(\zeta) (t) &:= -U_m^{\top} S \left( U_m \zeta(t)\right),\\
		\label{T}    \mathcal{T}(\zeta) &:= \eta_{lin}(t)  + A \zeta (t),
	\end{align}
	where $A:C \left( [0,T], \mathbb{R}^m  \right) \rightarrow C \left( [0,T], \mathbb{R}^m  \right)$ is a linear integral operator defined as
	\begin{equation}
		A \zeta (t) := \int_0^T L_m(t-s,T) \zeta(s) \mathrm{d}s.
		\label{A}
	\end{equation}
	Using the definitions in eqs.~\eqref{interpop}, \eqref{G}  and \eqref{T}, we can rewrite the collocation approximation eq.~\eqref{zzapprox} to the reformulated integral equation~\eqref{reforminteq} in a more compact manner as
	\begin{equation}
		\zeta^N = P^N \circ \mathcal{G} \circ \mathcal{T} (\zeta^N), \qquad \zeta^N \in V^N.
		\label{ziterative}
	\end{equation}
	With these preliminaries, the following theorem guarantees the convergence of our collocation solution $\zeta^N$ to eq.~\eqref{ziterative} and predicts the rate of convergence. 
	\begin{theorem}[Kumar \& Sloan \cite{kumar87}, Theorem 2]
		Let $\eta^* \in C \left( [0,T], \mathbb{R}^m  \right) $ be a geometrically isolated solution of eq.~\eqref{inteq}, and let $\zeta^*$ be the corresponding solution of eq.~\eqref{reforminteq}. Suppose the nonlinearity $S$ is of class $C^1$ and that the interpolatory operator $P^N$ is defined as in~\eqref{interpop}.
		Then:
		\begin{enumerate}[label=(\roman*)]
			\item There exists a natural number $N_0$ such that for $N \geq N_0$, eq.~\eqref{ziterative} has a solution $\zeta^N \in V^N$ satisfying
			\begin{equation}
			    \Vert \zeta^*-\zeta^N \Vert \rightarrow 0 \qquad \mathrm{as} \quad N \rightarrow \infty,
			    \label{thmy_firstassertion}
			\end{equation}
			and  $\eta^N=\mathcal{T}(\zeta^N)$ defines an approximation to $\eta^*$ satisfying
			$$
			\Vert \eta^*-\eta^N \Vert \rightarrow 0 \qquad \mathrm{as} \quad N \rightarrow \infty.
			$$
			\label{thm3_1}
			\item Suppose in addition that 1 is not an eigenvalue of the linear operator $D \left( \mathcal{G} \circ \mathcal{T} \right) (\zeta^*)$.
			Then there exists a neighborhood of $\zeta^*$ and a natural number $N_1$ such that for $N \geq N_1$ a solution $\zeta^N$ of \eqref{ziterative} is unique in that neighborhood. Furthermore, the approximation $\eta^N=\mathcal{T}(\zeta^N)$ satisfies 
			$$
			\Vert \eta^* - \eta^N \Vert \leq c \inf_{\phi \in V^N} \Vert \zeta^*-\phi \Vert,
			$$
			where $c>0 $ is independent of $N$.
			\label{thm3_2}
		\end{enumerate}
		\label{thmy}
	\end{theorem}
	\begin{proof}
		The proof carries over from Theorem~2 in~\cite{kumar87} with slight modifications. See Appendix \ref{appendix3} for details.
	\end{proof}
	
	Theorem~\ref{thmy} signifies that the collocation-based approximation~\eqref{zzapprox} of the reformulated integral equation~\eqref{reforminteq} converges to the corresponding solution of the original integral equation~\eqref{inteq}. Furthermore, the rate of convergence of $\eta^N$ to a geometrically isolated solution $\eta^{\star}$ of eq.~\eqref{inteq} is, at the very least, equal to the rate of convergence of the best approximation from $V^N$ to $\zeta^{\star}$, which is the corresponding solution of eq.~\eqref{reforminteq}.
	
	Computing the numerical solution to the nonlinear system~\eqref{zzapprox} for the unknown coefficients $b_i\in \mathbb{R}^m (i = 1,\dots , N)$,  involves the use of iterative methods, which we discuss next.

	\subsection{Iterative Methods}
	\label{iterativemethods}
	Starting with an initial guess $\zeta_0$, we are interested in obtaining an iterative approximation to a solution of the reformulated integral equation~\eqref{reforminteq} in the space of continuous, $T$-periodic functions. By Theorem~\ref{refromthm}, this solution will be in a one-to-one correspondence with a solution of the original integral equation~\eqref{inteq}, which in turn corresponds to the nonlinear periodic response of the dynamical system~\eqref{modelreduct} by Theorem~\ref{inteqthm}. 
	
	With the operator definitions~\eqref{G} and \eqref{T}, the original integral equation~\eqref{inteq} can be concisely written as
	\begin{equation}
		\eta = \mathcal{T}\circ \mathcal{G} (\eta),
		\label{TG}
	\end{equation}
	and its reformulated variant~\eqref{reforminteq} can be written as
	\begin{equation}
		\zeta = \mathcal{G} \circ \mathcal{T} (\zeta).
		\label{GT}
	\end{equation}
	We seek a solution to eq.~\eqref{GT} in a $\delta-$ball of continuous $T$-periodic functions centered around the initial guess $\zeta_0\in C \left([0,T],\mathbb{R}^m \right)$ defined as 
	\begin{equation}
		C_{\zeta_0,\delta}:=\left. \big\{ 
		\zeta \in C \left([0,T],\mathbb{R}^m \right) \; \right\vert \; \zeta(0)=\zeta(T), \quad \Vert \zeta - \zeta_0 \Vert_{\infty}\leq \delta  
		\big\}  .
		\label{C}
	\end{equation}
	To this end, we approximate the function $\zeta$ numerically within some finite-dimensional subspace of $ C_{\zeta_0,\delta}$. Specifically, using the definition~\eqref{V_N} of $V^N$, we define the collocation subspace
	\begin{equation}
		C_{\zeta_0,\delta}^N:=\left. \big\{ 
		\zeta^N \in V^N \; \right\vert \; \zeta^N(0)=\zeta^N(T), \quad \Vert \zeta^N - \zeta^N_0 \Vert_{\infty}\leq \delta  
		\big\} 
		\subset  C_{\zeta_0,\delta},
		\label{Cn}
	\end{equation}
	where $\zeta_0^N \in V^N$ with $\zeta_0^N(0) = \zeta_0^N(T)$ is an initial periodic solution guess in the collocation subspace. Finally, we aim to iteratively solve the system
	\begin{equation}
		\zeta^N =P^N \circ \mathcal{G} \circ \mathcal{T} (\zeta^N), \qquad \zeta^N \in C_{\zeta_0,\delta}^N,
		\label{approxiterativ}
	\end{equation}
	where $P^N:C_{\zeta_0,\delta} \rightarrow C_{\zeta_0,\delta}^N$ is the interpolatory projection operator defined in \eqref{interpop}.
	
	In the next section, we derive explicit conditions that ensure the convergence of the simple Picard iteration when applied to solve eq.~\eqref{approxiterativ}.
	
	\subsubsection{Picard Iteration}
	\label{picardsection}
	Note that solving eq.~\eqref{approxiterativ} is equivalent to obtaining fixed point(s) of the operator $P^N \circ \mathcal{G} \circ \mathcal{T}$. Jain, Breunung \& Haller~\cite{Jain2019} have already derived explicit conditions under which the simple Picard iteration, 
	\begin{equation}
		\eta_{\ell} = \mathcal{T}\circ \mathcal{G} (\eta_{\ell-1}), \quad \ell \in \mathbb{N},
	\end{equation}
	converges to a unique fixed point of the operator $\mathcal{T}\circ \mathcal{G}$ (see Theorem~5, Remark~9 in~\cite{Jain2019}). This provides an iterative solution of the original integral equation~\eqref{inteq}. However, their conditions are derived for the infinite-dimensional operator equation~\eqref{TG} and hence do not account for the collocation-approximation of the integral equation. Here, we derive similar estimates for the reformulated integral equation~\eqref{reforminteq}, taking into account the collocation approximation arising from the operator $P^N$. 
	
	First, we note that from Remark 2 in \cite{Jain2019}, we already have a formula for the operator norm of $ A $ (see eq.~\eqref{A}) as
	\begin{equation}
		\Vert A \Vert = \Gamma (T) :=  \max_{1\leq j \leq m} \frac{T \max \left( \vert e^{\lambda_j T} \vert,1 \right)}{\vert 1-e^{\lambda_j T} \vert}.
		\label{gamma}
	\end{equation}
	We further define the error of the first iteration step under the map $P^N \circ \mathcal{G} \circ \mathcal{T}$ as
	\begin{equation}
		\mathcal{E}(\zeta_0^N):= P^N\circ \mathcal{G} \circ \mathcal{T} (\zeta_0^N)-\zeta_0^N.
		\label{firstiterationerror}
	\end{equation}
	The following theorem establishes conditions under which the Picard iteration converges when applied to equation \eqref{approxiterativ}.
	
	\begin{theorem}
		Assume that the conditions
		\begin{enumerate}[label=(\roman*)]
			\item \begin{equation}
				\delta \geq \frac{\Vert \mathcal{E} (\zeta_0^N) \Vert_{\infty}}{1-\Vert P^N \Vert\Vert U_m^T \Vert L_S \Vert U_m \Vert \Gamma (T)}
				\label{cond1}
			\end{equation}  \label{picardass1}
			\item \begin{equation}
				L_S <\frac{1}{\Vert P^N \Vert\Vert U_m^T \Vert  \Vert U_m \Vert \Gamma (T)} 
				\label{cond2}
			\end{equation}  \label{picardass2}
		\end{enumerate}
		hold.
		Then the map $P^N\circ \mathcal{G} \circ \mathcal{T}$  has a unique fixed-point in the space $C_{\zeta_0,\delta}^N$ and this fixed point can be found via the convergent iteration 
		\begin{equation}
			\zeta_{\ell}^N=P^N \circ \mathcal{G} \circ \mathcal{T}(\zeta_{\ell-1}^N), \quad \ell\in \mathbb{N}.
			\label{picard}
		\end{equation}
		\label{picardthm}
	\end{theorem}
	
	\begin{proof}
		Analogous to the proof of Theorem 5 in \cite{Jain2019}, the proof here involves a direct application of the Banach fixed-point theorem, whereby assumptions \ref{picardass1} and \ref{picardass2} establish $P^N\circ \mathcal{G} \circ \mathcal{T}: C_{\zeta_0,\delta}^N \rightarrow C_{\zeta_0,\delta}^N   $ as a contraction. See Appendix~\ref{appendix4} for a detailed proof.
	\end{proof}
	
	The properties discussed in Section 3.1 of \cite{Jain2019} apply in the reformulated setting as well. Most notably, under the conditions~\eqref{cond1} and \eqref{cond2}, the Picard iteration converges monotonically to the unique periodic response in $C_{\zeta_0,\delta}^N$. Hence, an upper estimate for the error after a finite number of iterations is readily available as the sup norm of the difference $\Vert\zeta_{\ell}^N -\zeta_{\ell-1}^N \Vert_{\infty}$ of the last two iterations.
	
	We remark that when $m=n$ (i.e., no model reduction) and $\Vert P^N\Vert = 1$ (e.g., choosing orthonormal basis functions $v_i^N$for collocation), the estimates \ref{picardass1} and \ref{picardass2} for convergence of the Picard iteration~\eqref{picard} match those given in Theorem~5 of~\cite{Jain2019} for eq.~\eqref{TG}. As in~\cite{Jain2019}, we conclude from the dependence of \eqref{cond2} on the constant $\Gamma (T)$ that large damping, large separation of the forcing frequency $2\pi/T$ from the natural frequencies (i.e., larger
	$\vert 1-e^{\lambda_j T} \vert$) and high forcing frequencies (i.e., smaller $T$) are beneficial for the convergence of the Picard iteration.
	
	The use of the Picard iteration~\eqref{picard} specifically in the reformulated setting enables us to precompute the integral
	\begin{equation}
		\int_0^T L_m(t_j-s,T) v_i^N(s) \mathrm{d}s,
		\label{eq:Lint}
	\end{equation}
	which arises in eq.~\eqref{zzapprox}. As a result, the action of the operator $A$ (see eq~\eqref{A}) in the iteration~\eqref{zzapprox} becomes independent of $\zeta_{\ell-1}^N$ at the current iteration step. This leads to computational savings as discussed in Section~\ref{sectioncomparison}. We shall demonstrate these improvements on several numerical examples in Section~\ref{chapter4}.
	
	\subsubsection{Newton--Raphson iteration}
	\label{sec:NR}
	As noted in~\cite{Jain2019}, the convergence criteria for the Picard iteration will not be satisfied for near-resonant forcing and low damping. However, one or more periodic orbits might still exist even if the Picard iteration does not converge and the Newton--Raphson scheme provides a robust alternative in such cases. 
	An advantage of this iteration method is its quadratic convergence when the initial guess $\zeta_0^N$ is close enough to the actual solution.
	
	We now derive the Newton--Raphson scheme explicitly for equation \eqref{ziterative}.
	Let us define an operator $\mathcal{F}$ as 
	\begin{equation}
		\mathcal{F}(\zeta^N) := \zeta^N  - \mathcal{P}^N \circ\mathcal{G} \circ \mathcal{T} \left(\zeta^N\right).
		\label{Fdef}
	\end{equation}
	Then, looking for a fixed point of $ \mathcal{P}^N \circ\mathcal{G} \circ \mathcal{T}$ in eq.~\eqref{ziterative} is equivalent to solving
	\begin{equation}
		\mathcal{F} \left(\zeta^N\right) = 0.
		\label{Fzero}
	\end{equation}
	Starting with an initial guess $\zeta_0^N$, the classic Newton--Raphson iteration for eq.~\eqref{Fzero} is given as
	\begin{align}
		\zeta_{\ell+1}^N &= \zeta_{\ell}^N -[D \mathcal{F} (\zeta_{\ell}^N)]^{-1}\mathcal{F}(\zeta_{\ell}^N), \qquad \ell \in \mathbb{N},
		\label{eq:newtoniter}
	\end{align}
	where the Jacobian $D \mathcal{F} (\zeta_{\ell}^N)$ can be computed from eq.~\eqref{Fdef} as
	\begin{align}
		D \mathcal{F} (\zeta_{\ell}^N) \mu &= \mu - D (\mathcal{P}^N \circ\mathcal{G} \circ \mathcal{T} ) \left(\zeta_{\ell}^N\right) \mu\\
		&= \mu + P^N\left(U_m^{\top} D S \left(U_m \mathcal{T} \left(\zeta_{\ell}^N\right)\right) U_m A \mu\right).
		\label{eq:DF}
	\end{align}
	
	As for the Picard iteration (cf. Section~\ref{picardsection}), we can again precompute the action of the operator $A$ in the form of integrals such as \eqref{eq:Lint}. Hence, we expect computational savings from the reformulated integral equation (see Section~\ref{sectioncomparison}) under the Newton--Raphson iteration as well.
	
	The reformulated approach involves sparse matrix operations in the computation of the Jacobian~$D \mathcal{F}$, which leads to further speed gains relative to the original integral equation. Specifically, the action of $A$ can be written as a matrix, say $A_N$, constructed from blocks of diagonal matrices of the form \eqref{eq:Lint}. Now, in eq.~\eqref{Fdef}, the computation of the Jacobian $D \mathcal{F}$ involves the multiplication of the sparse matrix $A_N$ with another sparse matrix, which is the block-diagonalized version of the operator $U_m^T \circ D S (U_m \circ \mathcal{T} (\zeta_{\ell}^N)) \circ U_m$. On the other hand, in the original setting~\cite{Jain2019}, one must generally multiply a full matrix with a blockdiagonal matrix in order to obtain the corresponding Jacobian. This advantage is due to the specific reordering of operations introduced by the reformulation (cf eqs. \eqref{TG} and \eqref{GT}). 
	
	A drawback of the Newton-Raphson iteration scheme relative to the Picard iteration is that the Jacobian $D \mathcal{F} (\zeta_{\ell}^N)$ needs to be computed and inverted at each iteration step, which makes each iteration step much slower. At the same time, by its the quadratic convergence, the Newton--Raphson method requires much fewer iterations than the Picard iteration to converge. We will illustrate the benefits of these iterative methods on several numerical examples in the next section. 
	
	\section{Numerical examples}
	\label{chapter4}
	
	In this section, we demonstrate the performance gains obtained from the proposed reformulation of the original integral equations approach~\cite{Jain2019}. We also compare the speed gains arising from model reduction relative to the full system for these examples. For the latter, we employ the automated mode selection criterion developed in \cite{Buza2020} to obtain a projection-based ROM, as discussed in Section~\ref{modaltruncation}. 
	
	For these comparisons, we compute the forced response curves, i.e., the amplitude of the steady-state response as a function of the excitation frequency $\Omega =  2 \pi / T$ in a given range. We compute these curves via \emph{sequential continuation}, where we sweep through the range of excitation frequencies in discrete steps, using the converged solution of the previous (adjacent) step as the initial guess for the iteration at the current frequency step. Such an approach is generally bound to fail near a fold bifurcation with respect to the base frequency $\Omega$, where one can resort to more advanced continuation schemes such as the pseudo-arc-length continuation (see, e.g., \cite{Dankowicz2013a}). The simple sequential continuation is, however, sufficient for our exposition in this work. 
	
	For computing forced response curves, we optimize the benefits of the two iterative methods discussed above by using the Picard iteration away from resonances where it converges to the nonlinear periodic response very fast, and switching to the more robust Newton-Raphson method when Picard iteration fails to converge. We have included the implementation of our results along with the examples discussed below in an open-source MATLAB package~\cite{SST}.
	
	\subsection{Nonlinear oscillator chain}
	\label{sect:oscchainexample}
	We first consider an $n$-mass oscillator chain, which was used to demonstrate the computational performance of the original integral equations approach in \cite{Jain2019}.
	\begin{figure}[h]
		\includegraphics[width=0.6\textwidth]{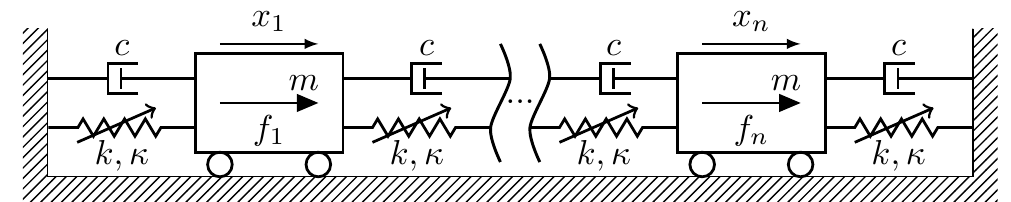}
		\centering
		\caption{An $n$-mass oscillator chain with coupled nonlinearities. In our analysis, we select the non-dimensional parameters $m=1$, $k=1$, $c=1$ and $\kappa=0.5$. Image taken from \cite{Jain2019}.}
		\label{fig:oscchain}
	\end{figure}
	The oscillator chain consists of $n$ oscillators of mass $m$ each, coupled with linear springs (with spring constant $k$), dampers (with damping coefficient $c$) and cubic springs (with coefficient $\kappa$), as shown in Figure~\ref{fig:oscchain}. The nonlinear function S (see eq.~\eqref{dynsys}) in this example is explicitly given as 
	\begin{equation}
		S(x) = \kappa
		\begin{pmatrix}
			x_1^3-(x_2-x_1)^3\\
			(x_2-x_1)^3-(x_3-x_2)^3\\
			\vdots \\
			(x_n-x_{n-1})^3-x_n^3\\
		\end{pmatrix}.    
	\end{equation}
	We consider a forcing of the form
	\begin{equation}
		f_i (t) = F \sin (\Omega t), \qquad i = 1,\ldots,n,    
	\end{equation}
	where $F$ denotes the forcing amplitude, such that each oscillator is excited with the same forcing at frequency~$\Omega$. For $n=20$, we first compare the results between the original and reformulated approach without any model reduction, i.e., $m=n$ in equations~\eqref{inteq} and \eqref{reforminteq}. Based on Theorem~\ref{refromthm}, we expect the same results from both approaches. Indeed, the corresponding forced response curves coincide at various forcing amplitudes, as shown in Figure~\ref{fig:oscchainresults}. Furthermore, we construct a ROM using the modal truncation~\eqref{modelreduct}, where $U_m$ comprises of the first 3 vibration modes ($m=3$), and compute the steady-state response for this ROM using the reformulated integral equation~\eqref{reforminteq}. Figure~\ref{fig:oscchainresults} further shows that this reduced periodic response accurately approximates the full system behavior.
	\begin{figure}[h]
		\includegraphics[width=0.5\textwidth]{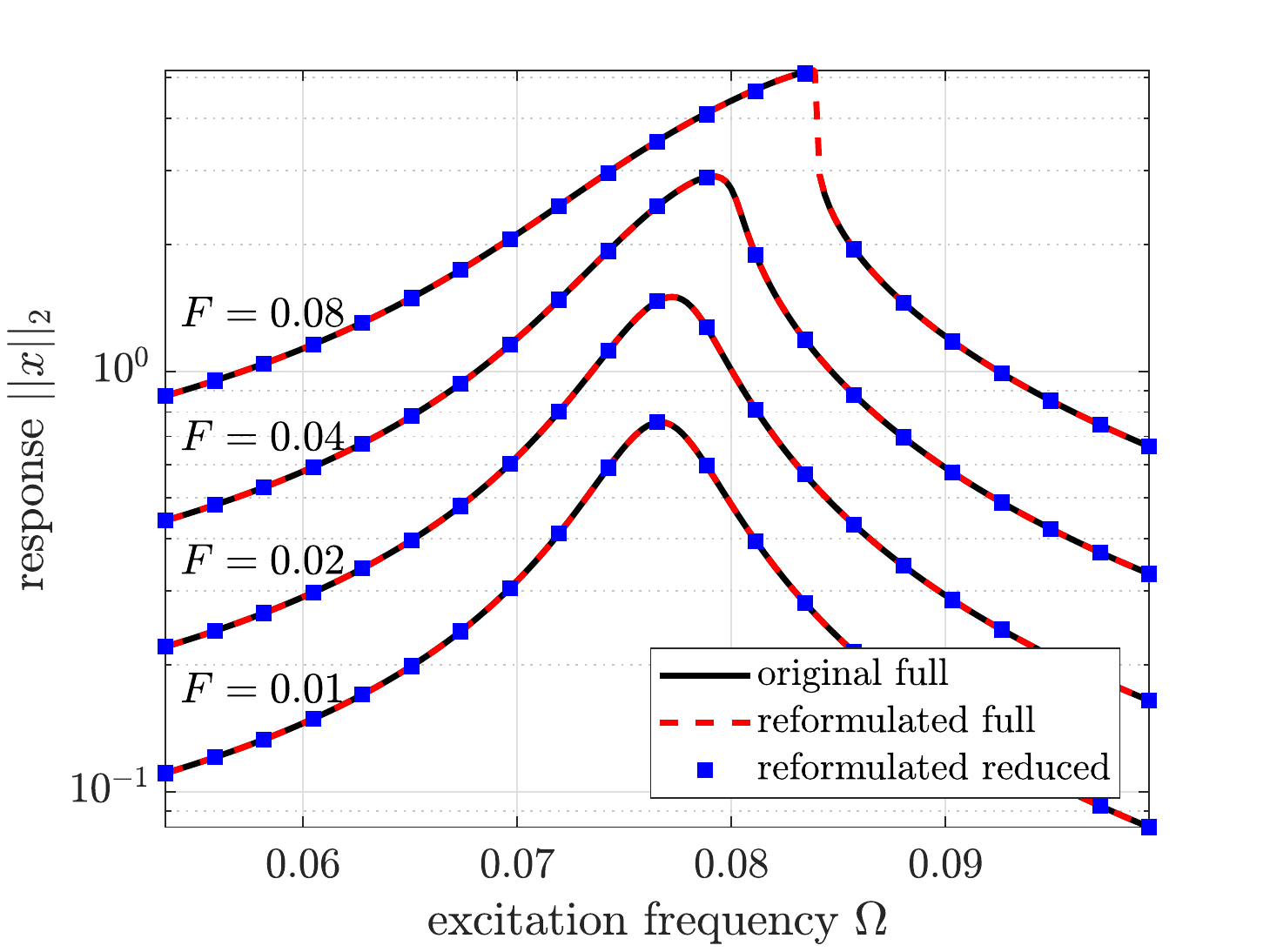}
		\centering
		\caption{\small The forced response curves obtained using the original and the reformulated approaches for various forcing amplitudes $F$ for the oscillator chain example (see Figure~\ref{fig:oscchain} with $n=20$ degrees of freedom.  As expected, the curves overlap when the two approaches are applied to the full system (i.e., $m = n$). Furthermore, the reduced periodic response obtained using the reformulated approach~\eqref{reforminteq} with the first three modes, accurately approximates the full periodic response.
		}
		\label{fig:oscchainresults}
	\end{figure}
	
	\begin{table}[h]
		\centering
		\sisetup{table-format=6.4}
		\begin{tabular}{lllll} \toprule
			{\multirow{3}{7em}{\textbf{Forcing amplitude} $F$}} & \multicolumn{4}{l}{ \textbf{Computation time} [seconds] (time spent on Picard, time spent on Newton--Raphson)} \\  \cmidrule{2-5}
			& {'original' full} & {'reformulated' full} & {'original' reduced} & {'reformulated' reduced} \\ 
			& {($m = n =20$)} & {($m = n =20$)} & {($m = 3$)} & {($m = 3$)} \\ 
			\toprule
			0.01  & 1.33 {(1.33,0)} &  0.83 {(0.83,0)} & 0.87 {(0.87,0)} & 0.71 {(0.71,0)} \\
			0.02  & 1.73 {(1.73,0)} & 1.05 {(1.05,0)} & 1.14 {(1.14,0)}  & 0.90 {(0.90,0)} \\
			0.04  & 7.77 {(2.09,5.68)} & 5.88 {(1.26,4.62)} & 2.19 {(1.42,0.77)}  & 1.83 {(1.16,0.67)} \\
			0.08 & 17.86 {(2.25,15.61)} & 11.91 {(1.35,10.56)} & 3.37 {(1.51,1.86)} & 2.76 {(1.19,1.57)} \\ \bottomrule
		\end{tabular}
		\caption{\small Computational time in seconds spent on obtaining the forced response curves of Figure~\ref{fig:oscchainresults} using the original~\eqref{inteq} and the reformulated~\eqref{reforminteq} integral equation approaches. The reformulated approach (columns 3  and 5) is consistently faster than the original approach (columns 2 and 4). The numbers in parentheses provides the break-down of the total computation time into contributions from the Picard and the Newton-Raphson iterations.}
		\label{tab:comparison2}
	\end{table}
	Table~\ref{tab:comparison2} records the computational time spent on obtaining the forced response curves of Figure~\ref{fig:oscchainresults} for different forcing amplitudes. We perform a sequential continuation to obtain these curves on a uniform grid of forcing frequency values for each entry in Table~\ref{tab:comparison2}. As mentioned earlier in this section, we use an optimal combination of Picard and Newton--Raphson iterations to obtain these curves. Along with the total computation time, Table~\ref{tab:comparison2} also records the part of time spent on the Picard and Newton--Raphson iterations separately in obtaining each of these curves. As expected, the 3-DOF ROM is an order of magnitude faster relative to the full system of 20 DOFs. Remarkably, we observe that the reformulated approach (see columns 3 and 5 in Table~\ref{tab:comparison2}) is consistently faster than the original integral equation (see columns 2 and 4 in Table~\ref{tab:comparison2}).
	
	\FloatBarrier
	\subsection{Curved von Kármán beam}
	\label{sect:curvedbeamexample}
	
	As a second example, we use a geometrically nonlinear von Kármán beam~\cite{JAIN2018195} with a curved geometry moving in a 2-dimensional plane. The curvature of the beam introduces a linear coupling between the axial and transverse degrees of freedom of the beam. As a consequence, various heuristic mode selection criteria become inapplicable, as discussed in \cite{Buza2020}.
	
	We follow a finite element discretization using cubic shape functions for the transverse displacements and linear shape functions for the axial displacements (cf.~\cite{JAIN2018195}). We assume a linear viscous damping model, which results in equations of motion of the general form~\eqref{dynsys} with purely position-dependent nonlinearities and proportional damping with 
	\begin{equation}
		C = \frac{\kappa}{E} K,
	\end{equation}
	where $E$ is the Young's modulus and $\kappa$ is the material damping coefficient.
	
	As in Section 6.2 of~\cite{Buza2020}, we consider an aluminium beam, discretized using 10 elements of equal size. The material parameters are $E = 70$ GPa, $\kappa = 0.1$ s$\cdot$GPa, $\rho = 2700$ kg/m$^3$ (density) and the geometric parameters $l = 1$ m (length), $h = 0.007$ m (height) and $b=0.1$ m (width). The curved beam is in the form of a circular arch such that its midpoint is raised by 5 mm relative to its ends. We choose doubly clamped boundary conditions at both ends of the beam, i.e., all degrees of freedom are constrained at both ends. We apply a uniform periodic forcing in the transverse direction, such that $f_i = F \sin (\Omega t)$, where the index $i$ represents each of the transverse displacement degrees of freedom.
	\begin{figure}[h]
		\includegraphics[width=0.5\textwidth]{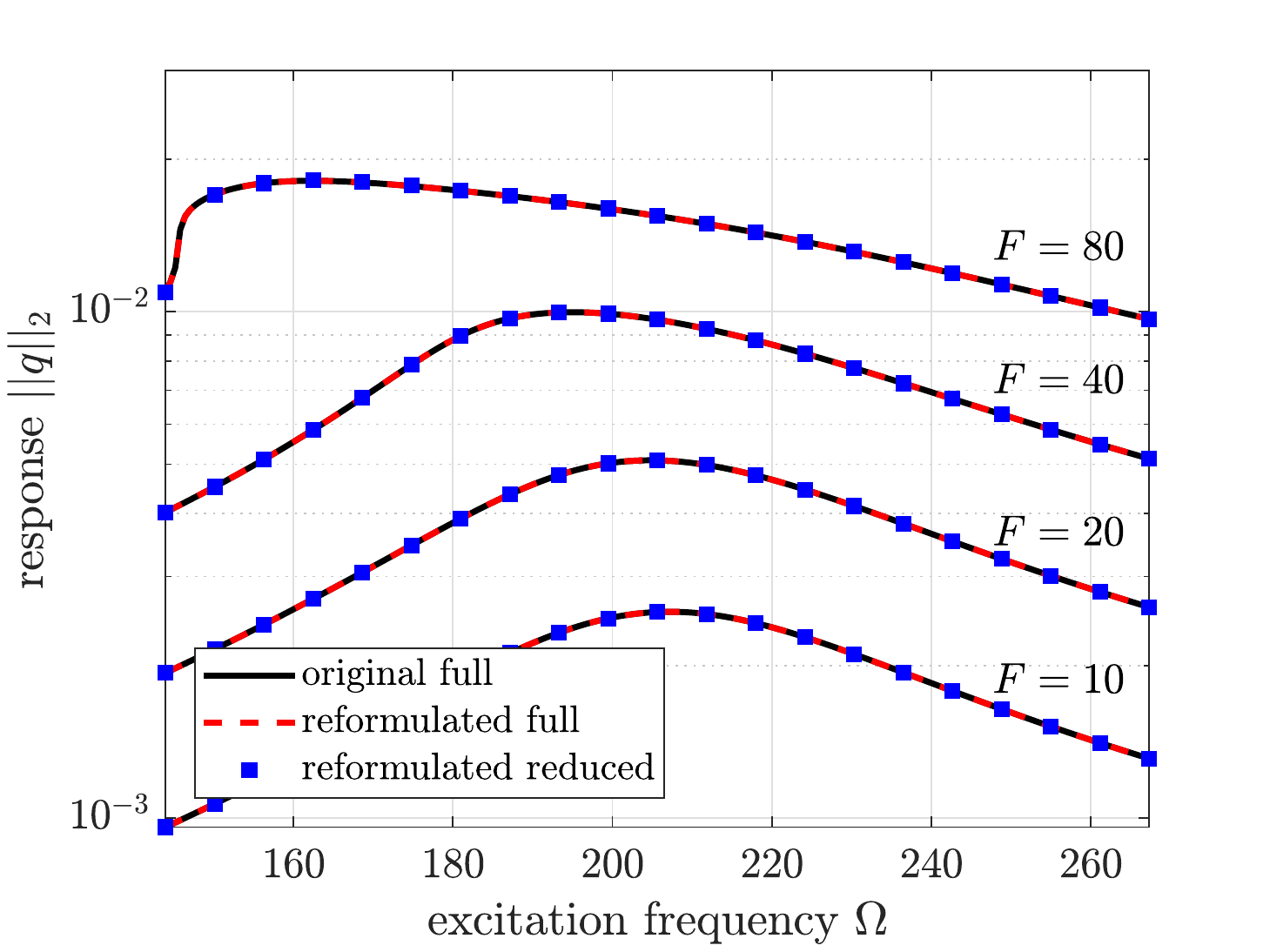}
		\centering
		\caption{\small The forced response curves obtained using the original and the reformulated approaches for various forcing amplitudes $F$ for the finite element model of a curved von Kármán with 10 elements.  As expected, the curves overlap when the two approaches are applied to the full system (i.e., $m = n$). Furthermore, the reduced periodic response obtained using the reformulated approach~\eqref{reforminteq} with the modes $ \{1,2,3,4,5,6,7,12,17 \}$~\cite{Buza2020} ($m=9$) accurately approximates the full periodic response.}
		\label{fig:curvedbeamresults}
	\end{figure}

	Similarly to the previous example (cf. Section~\ref{sect:oscchainexample}), we compare the forced response curves between the reformulated and original integral equations for different forcing amplitudes, as shown in Figure~\ref{fig:curvedbeamresults}. The ROM is constructed using the automated mode selection criterion in~\cite{Buza2020}, which returns the mode set $ \{1,2,3,4,5,6,7,12,17 \}$ ($m = 9$) for modal truncation. We observe that the reduced periodic response obtained using the reformulated approach~\eqref{reforminteq} via this optimal set of modes accurately approximates the full periodic response. Table~\ref{tab:comparison4} further compares the computation times for these forced response curves showing again consistently faster performance for the reformulated approach in both the full and the reduced setting.

	\begin{table}[h]
		\centering
		\sisetup{table-format=6.4}
		\begin{tabular}{lllll} \toprule
			{\multirow{3}{7em}{\textbf{Forcing amplitude} $F$}} & \multicolumn{4}{l}{ \textbf{Computation time} [seconds] (time spent on Picard, NR)} \\  \cmidrule{2-5}
			& {'original' full} & {'reformulated' full} & {'original' reduced} & {'reformulated' reduced} \\
			& {($m = n = 27$)} & {($m = n = 27$)} & {($m = 9$)} & {($m = 9$)}
			\\ \toprule
			10  & 5.12 {(5.12,0)} & 4.48 {(4.48,0)} & 1.72 {(1.72,0)}  & 1.50 {(1.50,0)} \\
			20  & 6.08 {(6.08,0)} & 5.23 {(5.23,0)} & 1.86 {(1.86,0)}  & 1.73 {(1.73,0)} \\
			40  & 23.76 {(7.44,16.32)} & 20.84 {(7.07,13.77)} & 3.67 {(2.32,1.35)} & 3.21 {(2.23,0.98)} \\
			80  & 77.88 {(2.33,75.55)} & 66.56 {(1.63,64.93)} & 6.87 {(0.75,6.12)} & 5.29 {(0.66,4.63)} \\ \bottomrule
		\end{tabular}
		\caption{\small Computational time in seconds spent on the forced response curves of Figure~\ref{fig:curvedbeamresults} using the original~\eqref{inteq} and the reformulated~\eqref{reforminteq} integral equation approaches. The reformulated approach (columns 3  and 5) is consistently faster than the original approach (columns 2 and 4). The numbers in parentheses provides the break-down of the total computation time into contributions from the Picard and the Newton-Raphson iterations.}
		\label{tab:comparison4}
	\end{table}

	\FloatBarrier
	\subsection{Curved plate model}
	\label{sect:curvedplate}
	
\begin{figure}[h]
		\includegraphics[width=0.5\textwidth]{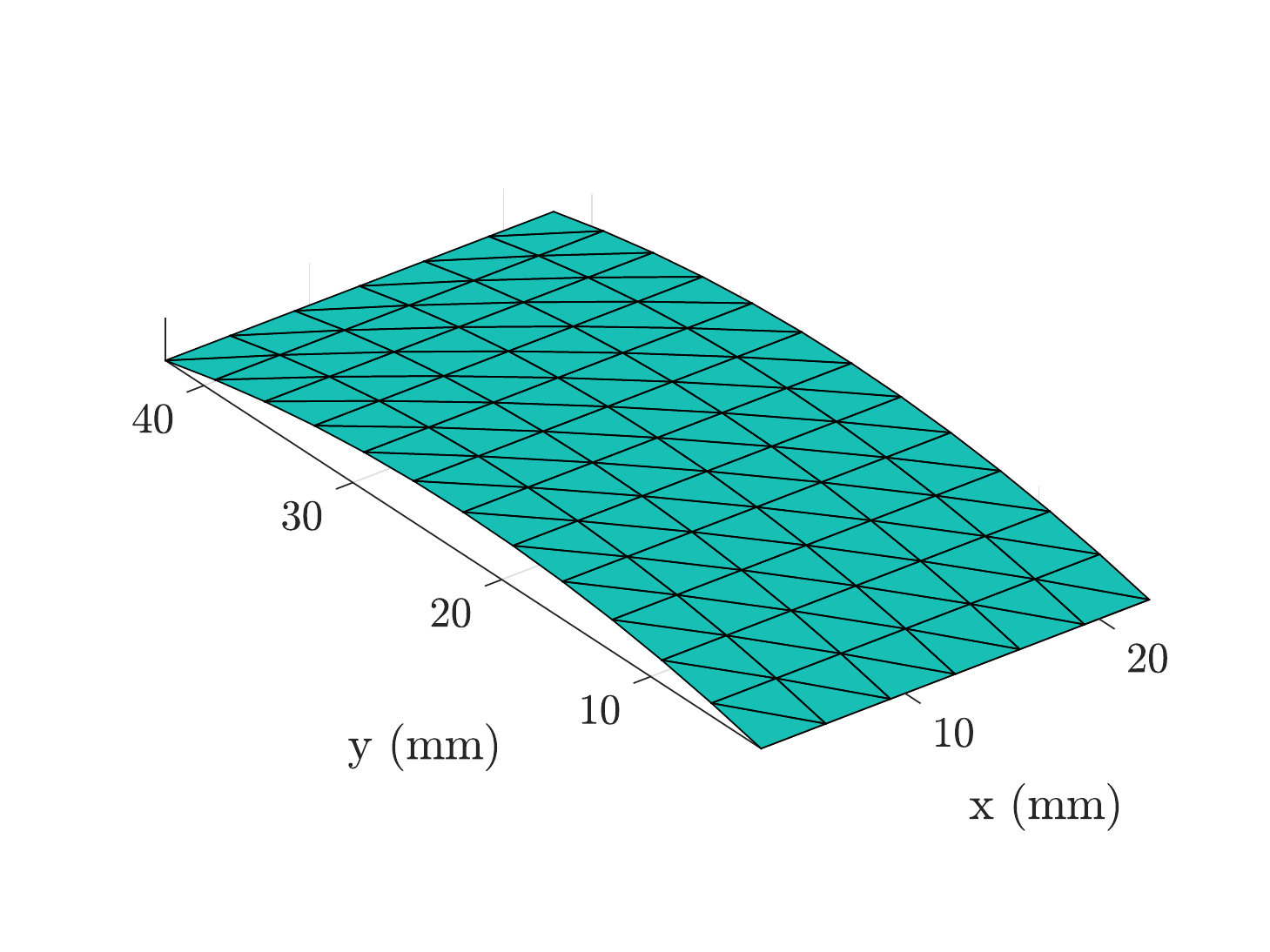}
		\centering
		\caption{\small Finite element mesh of a curved plate with length = 40mm, width = 20 mm and thickness = 0.8 mm. The curvature is cylindrical in nature along the y axis such that its midpoint is raised by $2$ mm relative to the short edges. We choose an aluminium material with parameters $E = 70$ GPa, $\rho = 2700$ kg/m$^3$ and simply supported boundary conditions on the shorter, opposite edges.}
		\label{fig:curvedplate}
	\end{figure}

	As a final example, we consider a shell-based finite element model of a curved, rectangular plate shown in Figure~\ref{fig:curvedplate}, which moves in a 3-dimensional space. As boundary conditions, we choose the two shorter, opposite edges of the plate to be simply supported, i.e., constrain the translational displacements along these edges in all directions. The mesh is generated using triangular shell elements with 6 degrees of freedom per node based on the von Kármán nonlinearities (see~\cite{10.1115/1.4040021,shellcode}).  The mesh constitutes $91$ nodes, which results in $504$ degrees of freedom after applying the boundary conditions.
	
	We consider an aluminium as material with parameters $E = 70$ GPa (Young's modulus), $\rho = 2700$ kg/m$^3$ (density) and the geometric parameters $l = 40$ mm (length), $b = 10$ mm (width), $h=0.8$ mm (thickness). The curved plate is in the form of a cylindrical arch such that its midpoint is raised by 2 mm relative to its shorter edges. We use Rayleigh damping with a modal damping factor of $4 \%$ for the first two modes. Once again, this results in governing equations~\eqref{dynsys} with purely geometric nonlinearities and proportional damping. We apply a uniform in space, periodic in time pressure on the top of the plate in the transverse direction, given in the form
	\begin{equation}
		p(t) = p_0 \sin (\Omega t).
	\end{equation}
	
\begin{figure}[h]
		\includegraphics[width=0.5\textwidth]{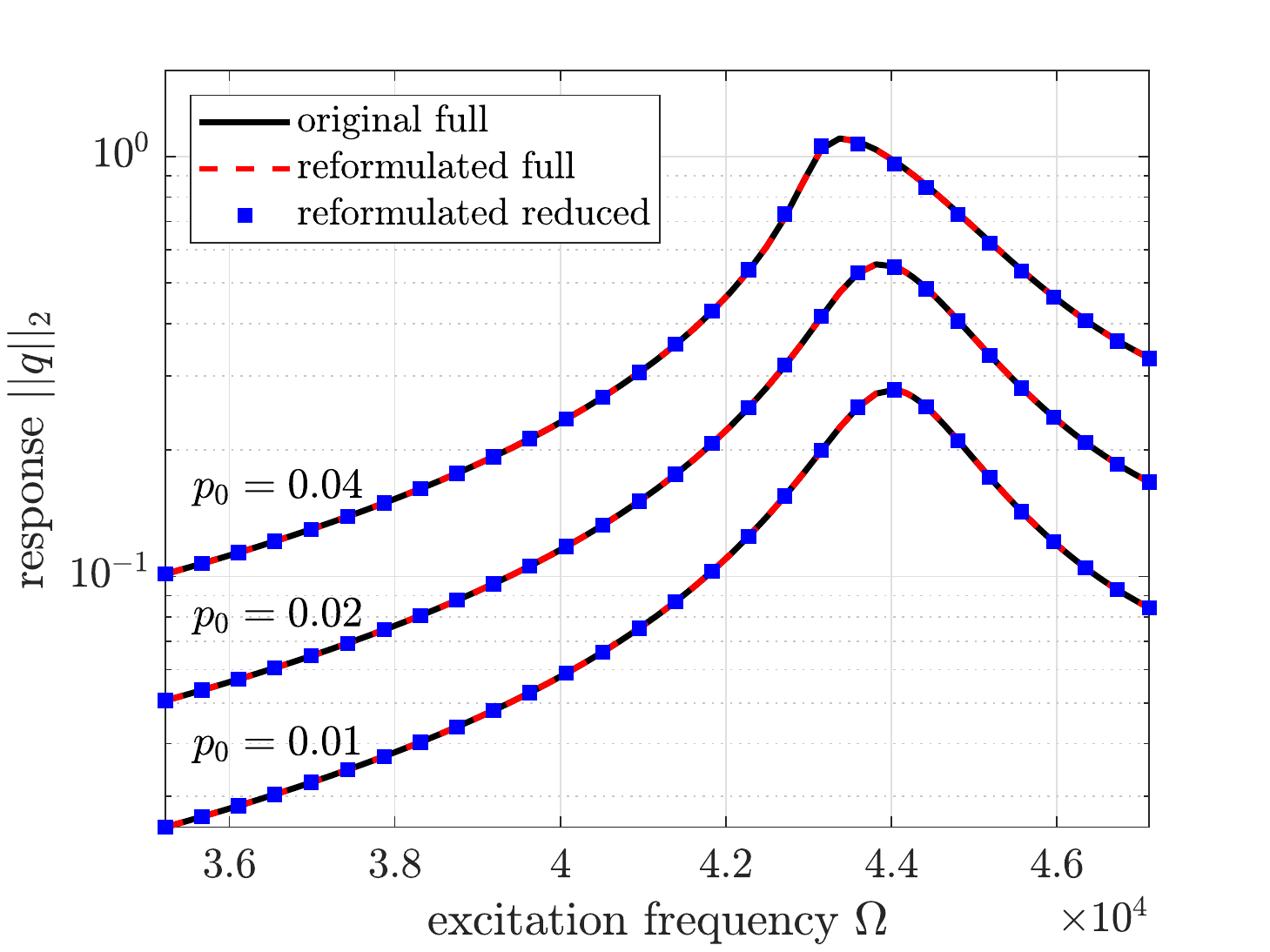}
		\centering
		\caption{\small The forced response curves obtained using the original and the reformulated approaches for various forcing amplitudes $p_0$ for the finite element model of a curved von Kármán plate with $n =504$ degrees of freedom. As expected, the curves overlap when the two approaches are applied to the full system (i.e., $m = n$). Furthermore, the reduced periodic response obtained using the reformulated approach~\eqref{reforminteq} with the modes $\{1,2,3,4,5,6,7,8,16,31 \}$~\cite{Buza2020} ($m = 10$), accurately approximates the full periodic response.
		}
		\label{fig:plateresults}
	\end{figure}

	For computing the forced response curves, we consider pressure amplitudes ranging from $p_0=0.01$ MPa up to $p_0 = 0.04$ MPa. We again use the the automated nonlinear mode selection procedure in \cite{Buza2020} for obtaining an optimal mode set for modal truncation, given by $\{1,2,3,4,5,6,7,8,16,31 \}$. The forced response curves showing softening type nonlinear behavior for different forcing amplitudes are depicted in Figure~\ref{fig:plateresults} and the corresponding computational times for different approaches are recorded in Table~\ref{tab:comparison6}. The results show trends analogous to the previous examples. The reformulation produces a faster computation of the same steady state response obtained using the original integral equation approach, and the automated mode-selection criterion~\cite{Buza2020} produces a reliable ROM for approximating the steady-state response using eq.~\eqref{reforminteq}. From the first two columns of Table \ref{tab:comparison6} we observe that the improvements on Newton-Raphson iteration steps are more significant for higher dimensional problems.
	Again, we discuss the reasons for this in the following section. 
	
	\begin{table}[h]
		\centering
		\sisetup{table-format=6.4}
		\begin{tabular}{lllll} \toprule
			{\multirow{3}{7em}{\textbf{Loading amplitude} ($p_0$)}} & \multicolumn{4}{l}{ \textbf{Computation time} [seconds] (time spent on Picard, NR)} \\  \cmidrule{2-5}
			& {'original' full} & {'reformulated' full} & {'original' reduced} & {'reformulated' reduced} \\
			& {($m = n = 504$)} & {($m = n = 504$)} & {($m = 10$)} & {($m = 10$)} 
			\\ \toprule
			0.01  & 1029 {(1029,0)} & 1028 {(1028,0)} & 917 {(917,0)}  & 917 {(917,0)} \\
			0.02  & 4877 {(949,3928)} & 2763 {(906,1857)} & 1176 {(910,266)}  & 1126 {(900,226)} \\
			0.04  & \textbf{11615} {(775,10840)} & \textbf{6411} {(770,5641)} & 1808 {(953,855)} & 1801 {(952,849)} \\ \bottomrule
		\end{tabular}
		\caption{\small Computational time in seconds spent in obtaining the forced response curves of Figure~\ref{fig:plateresults} using the original~\eqref{inteq} and the reformulated~\eqref{reforminteq} integral equation approaches. The reformulated approach (columns 3  and 5) is consistently faster that the original approach (columns 2 and 4). The numbers in parentheses provides the break-down of the total computation time into contributions from the Picard and the Newton-Raphson iterations.}
		\label{tab:comparison6}
	\end{table}
	
	The computation times spent on Picard steps are somewhat misleading in Table \ref{tab:comparison6}, since the most of these times is spent on evaluating the nonlinearities for these higher degree-of-freedom finite element systems. In fact, of the $1029$ seconds spent on computing the 'original full' response for the loading amplitude $p_0=0.01$ MPa, $1008$ seconds were contributed to nonlinear function evaluation. The relatively small differences in the Newton-Raphson steps of the reduced system can be attributed to the same fact because multiplying and inverting low dimensional matrices take significantly less time than the function evaluations. 
	\FloatBarrier
	\subsection{Discussion}
	\label{sect:observations}
	Tables \ref{tab:comparison2} and \ref{tab:comparison4} show that the Picard iteration method is consistently faster in the reformulated setting. As the Picard method takes the same number of iterations to converge in the reformulated and the original approaches (see Table~\ref{tab:comparison} for instance), the increase in speed arises from each iteration being faster. As discussed in Section \ref{picardsection},  this due to the fact that we can precompute convolution integrals of the type~\eqref{eq:Lint}. 
	
	\begin{table}[h]
		\centering
		\sisetup{table-format=6.4}
		\begin{tabular}{lllll} \toprule
			{\multirow{2}{7em}{\textbf{Forcing amplitude} $F$}} & \multicolumn{4}{l}{ \textbf{Total number of iterations}  (Picard, Newton--Raphson)} \\  \cmidrule{2-5}
			& {'original' full} & {'reformulated' full} & {'original' reduced} & {'reformulated' reduced} \\ \toprule
			0.01  & 733 {(733,0)} &  733 {(733,0)} & 734 {(734,0)} & 734 {(734,0)} \\
			0.02  & 967 {(967,0)} & 967 {(967,0)} & 967 {(967,0)}  & 967 {(967,0)} \\
			0.04  & 1240 {(1152,88)} & 1241 {(1152,89)} & 1240 {(1152,88)}  & 1241 {(1152,89)} \\
			0.08 & 1553 {(1286,267)} & 1496 {(1286,210)} & 1522 {(1296,226)} & 1522 {(1296,226)} \\ \bottomrule
		\end{tabular}
		\caption{\small Total number of iterations involved in obtaining the forced response curves of Figure~\ref{fig:oscchainresults} using the original~\eqref{inteq} and the reformulated~\eqref{reforminteq} integral equation approaches. The numbers in parentheses provides the break-down of the total number of iterations into contributions from the Picard and the Newton-Raphson method. The number of Picard iterations is the same for the reformulated approach (columns 3  and 5) as for the original approach (columns 2 and 4). }
		\label{tab:comparison}
	\end{table}
	
	We also observe that the Picard iteration converged for the same set of frequency values between the original and reformulated approaches in these examples. This is expected since the conditions for the convergence of Picard iteration in our reformulated setting (given by Theorem \ref{picardthm}) match the conditions for the original approach (given by Theorem 5 in~\cite{Jain2019}) as discussed in Section \ref{picardsection}.
	
	For the curved plate example, we observe that the computational gains from using the Picard iteration in the reformulated setting are marginal (see Table \ref{tab:comparison6}). This is because the primary computational bottleneck is the evaluation of the nonlinear function $S$, which is costly due to the finite element nature of the problem. This bottleneck may be alleviated by the use of hyperreduction methods (see, e.g.,~\cite{Farhat2014, 10.1115/1.4040021}, which aim at fast approximation of the nonlinearity by sampling the mesh. Such methods can also be equipped with our proposed reformulation of the integral equations approach. 
	
	The Newton--Raphson iteration is also consistently faster in the reformulated setting (see Table~\ref{tab:comparison6} in particular). Aside from the precomputation of the integral, this speed gain results from the sparsity operators arising in the computation of the Jacobian~$D\mathcal{F}$ (see eq.~\eqref{eq:DF}) in the reformulated setting, as discussed in Section~\ref{sec:NR}.

	\section{Conclusions}
	
	Jain, Breunung and Haller~\cite{Jain2019} proposed an integral equation for the fast computation of the steady-state response of nonlinear mechanical systems under (quasi)periodic forcing. In this work, we have proposed a reformulation to this integral equation based on the results of Kumar and Sloan \cite{kumar87}, which leads to better computational performance. We have established the one-to-one correspondence between solutions of the original and reformulated integral equations (Theorem~\ref{refromthm}) for which, we have extended the scalar results of Kumar and Sloan~\cite{kumar87} for vector-valued functions (Appendix~\ref{appendix2}). 
	
	We have performed numerical analysis of the reformulated approach and have discussed the Picard and Newton--Raphson iterative methods to solve these integral equations in Section~\ref{sectionapprox}. We conclude that the proposed reformulation leads to categorically better computational performance relative to the original integral equation approach in~\cite{Jain2019} when using the Picard or the Newton--Raphson iterations. We have used an optimal combination of the Picard and the Newton--Raphson iterations to compute the forced response curves in periodically forced mechanical systems. We have derived explicit conditions that guarantee convergence of the fast Picard iteration (Theorem~\ref{picardthm}). In contrast, we have used the more robust but expensive Newton--Raphson method when the Picard iteration failed to converge. 
	
	Finally, we have integrated this approach with modal truncation-based reduced-order modeling using the automated mode selection procedure developed in ref.~\cite{Buza2020}. We have demonstrated the gains in computational performance from the proposed reformulation on several numerical examples of varying complexity in Section~\ref{chapter4} both with and without modal truncation. We observe that modal truncation using the optimal mode selection criterion results in a significant increase in the computational performance with negligible losses in accuracy, as apparent from all examples (cf. Tables \ref{tab:comparison2}, \ref{tab:comparison4} and \ref{tab:comparison6} with Figures \ref{fig:oscchainresults}, \ref{fig:curvedbeamresults} and \ref{fig:plateresults}). An open-source implementation~\cite{SST} of the results is now available which serves as an update to the original \textsc{MATLAB} code of Jain, Breunung and Haller~\cite{Jain2019}.
	
	\appendix
	
	\section{Extension of Kumar \& Sloan's results~\cite{kumar87} to vector-valued functions}
	\label{appendix2}
	
	\subsection{Function spaces and preliminary definitions}
	\label{appendix1}
	
	For any two normed vector spaces $V$ and $W$, we denote by $\mathcal{L}(V,W)$ the space of continuous (i.e., bounded) linear operators from $V$ to $W$ equipped with the operator norm
	$$
	\Vert A \Vert_{\mathcal{L}(V,W)}  = \sup_{v \in V, \Vert v \Vert \leq 1} \Vert A v \Vert.
	$$
	Next, we recall the notion of compact and completely continuous operators (see \cite{kantorovich1982functional}, for instance), which are frequently used in the subsequent proofs.
	\begin{definition}
		A bounded operator $A \in \mathcal{L}(V,W)$ is said to be \emph{compact} if $A(B_1(0))$ has compact closure in $W$, where $B_1(0)$ denotes the unit ball in $V$.
		\label{compactdef}
	\end{definition}
	\begin{definition}
		Let $V$ and $W$ be Banach spaces.
		An operator $A:V \rightarrow W$ is said to be \emph{completely continuous} if it is continuous and maps any bounded subset of $V$ into a relatively compact subset of $W$.
		\label{ccdef}
	\end{definition}
	In this work, we are concerned with the  space of continuous functions defined on a closed interval $\left[ a, b \right]$ that take values in $\mathbb{R}^n $, i.e.,
	$$
	C \left( \left[ a,b \right], \mathbb{R}^n \right) := \left.
	\big\{ f : \left[ a,b \right] \rightarrow \mathbb{R}^n \; \right\vert \; f \text{ is continuous} \big\},
	$$
	We equip this vector space with the supremum norm
	\begin{equation}
		\Vert f \Vert_{\infty} = \sup_{a \leq t \leq b} \Vert f(t) \Vert_2, \qquad f \in C \left( \left[ a,b \right], \mathbb{R}^n \right), 
		\label{norm}
	\end{equation}
	which makes $ \left( C \left( \left[ a,b \right], \mathbb{R}^n \right), \Vert \cdot \Vert_{\infty} \right)$ a Banach space.
	
	\subsection{Integral equations}
	We consider general, vector-valued integral equations of the form
	\begin{equation}
		y(t) = f(t) + \int_a^b K(t,s) g \left( s,y \left(s\right) \right) \mathrm{d}s, \qquad t \in \left[ a, b \right],
		\label{hammerstein1}
	\end{equation}
	where $-\infty <a<b< \infty$, $f:\left[ a, b \right] 	\rightarrow \mathbb{R}^n $,
	$K: \left[ a, b \right] \times \left[ a, b \right] \rightarrow \mathbb{R}^{n\times n} $,
	and $g: \left[ a, b \right] \times \mathbb{R}^n \rightarrow \mathbb{R}^n$ are known functions
	and $y$ is the solution to be determined. The function $g$ is assumed to be nonlinear in its second argument.
	
	Following Kumar \& Sloan~\cite{kumar87}, we define a new function $z:\left[ a, b \right] \rightarrow \mathbb{R}^n $ by
	$
	z(t) := g\left( t,y \left(t\right) \right).
	$
	In the following, we show that
	equation \eqref{hammerstein1} is equivalent to a reformulated equation of the form
	\begin{equation}
		z(t) = g \left( t, f(t) + \int_a^b K(t,s) z(s) \, \mathrm{d}s  \right), \qquad t \in \left[ a, b \right],
		\label{reformulated}
	\end{equation}
	i.e., the solutions of eq.~\eqref{reformulated} are in bijective correspondence with solutions of \eqref{hammerstein1}.
	
	We further define the assumptions on the functions $f$, $g,$ and $K$ analogous to Kumar \& Sloan~\cite{kumar87}
	\begin{enumerate}[label=(A\arabic*)]
		\item
		$
		\sup_{a \leq t \leq b}    \int_a^b \Vert K(t,s) \Vert_2 \, \mathrm{d}s  < \infty,
		$
		\label{ass1}
		\item
		$
		\lim_{t \rightarrow t'} \int_a^b \left\Vert K(t,s)-K(t',s) \right\Vert_2 \mathrm{d}s= 0, \qquad t' \in \left[ a, b \right],
		$
		\label{ass2}
		\item $ f \in C \left( \left[ a,b \right], \mathbb{R}^n \right)$, \label{ass3}
		\item the function $g$  is defined and continuous on $[a,b] \times \mathbb{R}^n $, \label{ass4}
		\item the function $\partial g / \partial v$ is defined and continuous on $[a,b] \times \mathbb{R}^n $. \label{ass5}
	\end{enumerate}
	
	Next, we define some operators related to the integral equations~\eqref{hammerstein1} and \eqref{reformulated}. Let $A: C \left( \left[ a,b \right], \mathbb{R}^n \right) \rightarrow
	C \left( \left[ a,b \right], \mathbb{R}^n \right)$ denote the linear integral operator defined as
	\begin{equation}
		(Aw) (t) : = \int_a^b K(t,s)w(s) \, \mathrm{d}s, \qquad t \in \left[ a,b \right].
		\label{Adef}
	\end{equation}
	Let $\mathcal{T}$ be the nonlinear operator defined by
	\begin{equation}
		\mathcal{T}(w) := f + (Aw) ,  \qquad w \in  
		C \left( \left[ a,b \right], \mathbb{R}^n \right).
		\label{eq:T}
	\end{equation}
	Finally, we define a substitution operator for the function $g$ by
	\begin{equation}
		\mathcal{G}(y) (t) := g(t,y(t)), \qquad t \in [a,b], \quad y \in 
		C \left( \left[ a,b \right], \mathbb{R}^n \right).
		\label{substitution}
	\end{equation}
	Equations \eqref{hammerstein1} and \eqref{reformulated} can now be written in operator form as
	\begin{equation}
		y = \mathcal{T} \circ \mathcal{G} (y) , \qquad y \in C \left( \left[ a,b \right], \mathbb{R}^n \right),
		\label{operatory}
	\end{equation}
	and
	\begin{equation}
		z = \mathcal{G} \circ  \mathcal{T} (z) , \qquad z \in C \left( \left[ a,b \right], \mathbb{R}^n \right).
		\label{operatorz}
	\end{equation}
	
	\subsection{One-to-one correspondence between solutions of eqs.~\eqref{hammerstein1} and \eqref{reformulated}}
	
	With the operator equations~\eqref{operatory} and \eqref{operatorz} at hand, we can recall a result that has been pointed out by \cite{krasnosel1984geometrical} (see page 143) in the form of the following lemma. 
	\begin{lemma}[Kumar \& Sloan \cite{kumar87}, Lemma 1] The operator $\mathcal{G}$ is a bijection from the solution set $\Theta_{ \mathcal{T} \circ \mathcal{G} }:=
		\left\{ y \in C \left( \left[ a,b \right], \mathbb{R}^n \right) \; \vert \; 
		\mathcal{T} \circ \mathcal{G}(y) = y \right\}$
		onto
		$\Theta_{ \mathcal{G} \circ \mathcal{T} }:=
		\left\{ z \in C \left( \left[ a,b \right], \mathbb{R}^n \right) \; \vert \; 
		\mathcal{G} \circ \mathcal{T}(z) = z \right\}$, with inverse $\mathcal{T}$.
		\label{kumarsloan}
	\end{lemma}
	
	This guarantees the one-to-one correspondence between solutions of \eqref{hammerstein1} and \eqref{reformulated}.
	
	\subsection{Compactness of $A$}

	Establishing the compactness of $A$ is a vital step in confirming the convergence result stated in Theorem \ref{thmy}. Proving compactness in the vector-valued case requires a more generalized formulation of the Arzelà–Ascoli theorem (see \cite{buhler2018functional}) than the one used by Kumar and Sloan \cite{kumar87}. We first recall the notion of equicontinuity used in the theorem.
	\begin{definition}
	\label{equicont}
		Let $(X,d_X)$ and $(Y,d_Y)$ be two metric spaces. A family of functions $\mathscr{F} \subset C(X,Y)$ is called \textit{equicontinuous} if $ \forall \varepsilon>0$ $ \exists \delta>0$ such that $ \forall x,x' \in X$ and $\forall f \in \mathscr{F}:$ $d_X(x,x') < \delta \implies d_Y(f(x),f(x'))< \varepsilon$.
	\end{definition}
	
	\begin{theorem}[Arzelà–Ascoli]
		Let $(X,d_X)$ be a compact metric space and $(Y,d_Y)$ be a complete metric space. Consider a subset $\mathscr{F} \subset C(X,Y)$. \newline
		The following are equivalent:
		\begin{enumerate}[label=(\roman*)]
			\item $\mathscr{F}$ has a compact closure.
			\item $\mathscr{F}$ is equicontinuous and $\mathscr{F}(x):=\left\{ f(x) \; \vert \; f \in \mathscr{F} \right\} \subset Y$ has a compact closure for every $x \in X$.
		\end{enumerate}
		\label{arzela-ascoli}
	\end{theorem}
	
	With these preliminaries, we prove the compactness of $A$ in the following statement.  
	\begin{lemma}
		Let $A: C \left( \left[ a,b \right], \mathbb{R}^n \right) \rightarrow
		C \left( \left[ a,b \right], \mathbb{R}^n \right)$ be the linear integral operator defined in~\eqref{Adef}. Then under the assumptions \ref{ass1} and \ref{ass2}, $A$ is a compact operator.
		\label{lemmaA2}
	\end{lemma}
	\begin{proof}
		Let $B_1(0)$ denote the unit ball in $C \left( \left[ a,b \right], \mathbb{R}^n \right)$ defined as
		$$
		B_1(0) := \left. \big\{ w \in C \left( \left[ a,b \right], \mathbb{R}^n \right) \;
		\right\vert \; \Vert w \Vert_{\infty} < 1 \big\}.
		$$
		First we show that $A \left( B_1(0) \right)$  is equicontinuous.
		Fix $\varepsilon>0$.
		Then for any $w \in B_1(0) $
		$$
		\Vert (Aw)(t)-(Aw)(t') \Vert_2 = \left\Vert \int_a^b \left( K(t,s) - K(t',s) \right) w(s) \, \mathrm{d}s \right\Vert_2 \leq \int_a^b \left\Vert K(t,s)-K(t',s) \right\Vert_2 \mathrm{d}s.
		$$
		By assumption \ref{ass2}, for all $t' \in [a,b]$ we can choose $\delta>0$ such that for $\vert t - t' \vert < \delta$, $\int_a^b \left\Vert K(t,s)-K(t',s) \right\Vert_2 \mathrm{d}s< \varepsilon$. This shows pointwise equicontinuity at all $t' \in [a,b]$, which implies uniform equicontinuity in the sense of Definition \ref{equicont} by compactness of $[a,b]$. \newline
		Next we show that $A \left( B_1(0) \right)(t)$ has compact closure for every $t \in \left[ a,b \right]$.
		Pick $w \in B_1(0) $ and $t \in [a,b]$.
		Then
		$$
		\Vert (Aw) (t) \Vert_2 \leq \int_a^b \Vert K(t,s) \Vert_2 \, \mathrm{d}s \leq \sup_{a \leq u \leq b}    \int_a^b \Vert K(u,s) \Vert_2 \, \mathrm{d}s  < \infty,
		$$
		by assumption \ref{ass1}. Since $w$ and $t$ were arbitrary, $A \left( B_1(0) \right) (t)$ is bounded for all $t \in [a,b]$.
		Clearly $\overline{A \left( B_1(0) \right)(t)}$ is bounded whenever $A \left( B_1(0) \right)(t)$ is bounded.
		By the Heine-Borel theorem a subset of $\mathbb{R}^n $ is compact if and only if it is closed and bounded. We can now apply Theorem~\ref{arzela-ascoli} with $\mathscr{F} = A \left( B_1(0) \right)$, $X=[a,b]$, and $Y = \mathbb{R}^n $ to conclude that $A(B_1(0))$ has compact closure.
	\end{proof}
	
    Being compact and linear, $A$ is necessarily completely continuous (see. \cite{kantorovich1982functional}, p. 244). Furthermore, $\mathcal{T}$ is also a completely continuous operator like $A$, as it differs from $A$ only due to the inhomogeneous term.
	
	The substitution operator \eqref{substitution} is continuous and bounded if $g$ is continuous in both variables (see Section 17.7 in \cite{krasnosel1984geometrical}) (assumption~\ref{ass4}). Since $\mathcal{T}$ is completely continuous and $\mathcal{G}$ is continuous and bounded, it is clear from Definition \ref{ccdef} that $\mathcal{T} \circ \mathcal{G}$ is completely continuous. The complete continuity of $\mathcal{G} \circ \mathcal{T}$ is immediate from the continuity of $\mathcal{G}$, since continuous operators map compact sets into compact sets.
	
	Once the complete continuity of operators $\mathcal{T} \circ \mathcal{G}$ and $\mathcal{G} \circ \mathcal{T}$ is established, we are essentially in the setting of Kumar and Sloan \cite{kumar87}, and all the remaining results follow with some slight notational alterations.
	We repeat these results here for the sake of completeness.
	
	\subsection{Numerical analysis of the collocation approximation}
	\label{sec:KumarSloancollocation}
	
	For an analysis of the numerical method we consider \textit{geometrically isolated} solutions of
	\eqref{operatory} \cite{geomiso}.
	A solution $y^*$ of \eqref{operatory} is \textit{geometrically isolated} if it is the only solution of 
	\eqref{operatory} in some ball centered at $y^*$.
	
	\begin{lemma}[Kumar \& Sloan \cite{kumar87}, Lemma 2]
		Suppose assumptions \ref{ass1} to \ref{ass4} hold.
		If $y^*$ is a geometrically isolated solution of \eqref{operatory}, then $z^*:=\mathcal{G}(y^*)$ is
		a geometrically isolated solution of \eqref{operatorz}.
		Conversely, if $z^*$ is a geometrically isolated solution of \eqref{operatorz}, then 
		$y^*:=\mathcal{T}(z^*)$ is a geometrically isolated solution of \eqref{operatory}.
		\label{geomisollemma}
	\end{lemma}
	\begin{proof}

		The proof follows from the continuity of $\mathcal{G}$ and $\mathcal{T}$.
	\end{proof}
	
	Subsequent results of Kumar and Sloan \cite{kumar87} make use of a topological approach to nonlinear equations.
	For this, the notion of the index of a singular point is used, which is analogous to the notion of multiplicity for zeroes of polynomials in algebra.
	The index is defined via the common value of the \textit{rotation} of a vector field around an isolated singular point.
	For a proper definition of the rotation and a better understanding of these concepts, we refer to pages 5-9 in \cite{krasnosel1984geometrical}.
	By these definitions, the index of a geometrically isolated solution $y^*$ is the common value of rotation of
	$
	\mathrm{id} - \mathcal{T} \circ \mathcal{G} 
	$
	over all sufficiently small spheres centered at $y^*$, where $
	\mathrm{id}: C \left( \left[ a,b \right], \mathbb{R}^n \right) \rightarrow
	C \left( \left[ a,b \right], \mathbb{R}^n \right) 
	$ denotes the identity map.
	
	\begin{lemma}[Kumar \& Sloan \cite{kumar87}, Lemma 3]
		Suppose assumptions \ref{ass1} to \ref{ass4} hold.
		Let $y^*$ be a geometrically isolated solution of \eqref{operatory} and let $z^*$ be the corresponding geometrically isolated solution of \eqref{operatorz}. Then $y^*$ and $z^*$ have the same index.
	\end{lemma}
	\begin{proof}
		This is a special case of Theorem 26.3 of \cite{krasnosel1984geometrical}
		(this theorem uses that $\mathcal{G} \circ \mathcal{T}$ and
		$\mathcal{T} \circ \mathcal{G}$ are completely continuous).
	\end{proof}
	
	The following lemma establishes conditions under which the operator $\mathcal{G} \circ \mathcal{T}$ is
	Fréchet differentiable.
	
	\begin{lemma}[Kumar \& Sloan \cite{kumar87}, Lemma 4]
		Suppose assumptions  \ref{ass1} to  \ref{ass5} hold.
		Then $\mathcal{G}$ is continuously Fréchet differentiable on $C \left( \left[ a,b \right], \mathbb{R}^n \right)$; its Fréchet derivative at $x \in C \left( \left[ a,b \right], \mathbb{R}^n \right)$ is the multiplicative linear operator $D \mathcal{G} (x)$ given by
		\begin{equation}
			D\mathcal{G}(x) w  (t)  = \left. \frac{\partial g}{\partial v} \right|_{ (t,x(t))}w(t), \qquad
			t \in [a,b], \quad w \in C \left( \left[ a,b \right], \mathbb{R}^n \right).
			\label{frechg}
		\end{equation}
		Furthermore $\mathcal{G} \circ \mathcal{T}$ is continuously Fréchet differentiable on $C \left( \left[ a,b \right], \mathbb{R}^n \right)$; its Fréchet derivative at $x \in C \left( \left[ a,b \right], \mathbb{R}^n \right)$ is the completely continuous linear operator $D \left( \mathcal{G} \circ \mathcal{T} \right)(x)$ given by
		\begin{equation}
			D \left( \mathcal{G} \circ \mathcal{T} \right)(x) w  (t) =
			\left. \frac{\partial g}{ \partial v} \right|_{ \left( t,\mathcal{T}(x)(t) \right)}Aw(t),
			\qquad t \in [a,b], \quad w \in C \left( \left[ a,b \right], \mathbb{R}^n \right).
			\label{frechgt}
		\end{equation}
		\label{frechetlemma}
	\end{lemma}
	\begin{proof}
		The first result \eqref{frechg} is obvious, the second result \eqref{frechgt} uses the chain rule:
		$$
		D \left( \mathcal{G} \circ \mathcal{T} \right)(x) w  (t) = 
		D  \mathcal{G} (\mathcal{T} (x)) \circ D \mathcal{T} (x) w  (t) \overset{(\dagger)}{=}
		\left. \frac{\partial g}{ \partial v} \right|_{ \left(t,\mathcal{T}(x)(t) \right)}Aw(t),
		$$
		where the equality ($\dagger$) follows from $D \mathcal{T} (x)  = A$. The fact that $D \left( \mathcal{G} \circ \mathcal{T} \right)(x)$ is completely continuous is a direct consequence of Theorem 17.1 in \cite{krasnosel1984geometrical}, which uses the fact that $\mathcal{G} \circ \mathcal{T}$ itself is completely continuous.
	\end{proof}
	
	\begin{lemma}[Kumar \& Sloan \cite{kumar87}, Proposition 1]
		Assume the conditions \ref{ass1} to  \ref{ass5} hold, and let $z^*$ be a solution of
		\eqref{operatorz}. If 1 is not an eigenvalue of the linear operator $D \left( \mathcal{G} \circ \mathcal{T} \right)(z^*)$, then $z^*$ is geometrically isolated and has index $\pm 1$.
		\label{lemma19}
	\end{lemma}
	\begin{proof}
		This is Theorem 21.6 in \cite{krasnosel1984geometrical}.
	\end{proof}
	
	From the definition~\eqref{interpop}, we see that $P^N$ is a bounded linear operator with norm
	$$
	\Vert P^N \Vert = \sup_{0 \leq t \leq T} \sum_{i=1}^N \left\vert v^N_i(t) \right\vert,
	$$
	and that $P^N \circ P^N = P^N$. 
	Assume that the subspace $V^N$ and the collocation points $\{t_i\}_{i=1}^N$ are such that
	\begin{equation}
		\lim_{N \rightarrow \infty} \Vert w-P^N w \Vert = 0, \qquad \text{for all} \quad w \in C \left( [0,T], \mathbb{R}^n  \right).
		\label{thmreq}
	\end{equation}
	Then, by the uniform boundedness principle (Banach–Steinhaus), there exists $c>0$ such that $\Vert P^N \Vert \leq c$ for all $N$, i.e., $P^N$ is uniformly bounded as an operator from $C \left( [a,b], \mathbb{R}^n  \right)$ to $V^N$. 
	
	The next result is a direct application of Theorem 19.7 of \cite{krasnosel1972approximate} along with the results of Lemma \ref{lemmaA2} - \ref{lemma19}.

	\begin{theorem}[Kumar \& Sloan \cite{kumar87}, Theorem 1]
		Let $y^* \in C \left( [a,b], \mathbb{R}^n  \right) $ be a geometrically isolated solution of \eqref{operatory}, and let $z^*$ be the corresponding solution of \eqref{operatorz}.
		Suppose conditions \ref{ass1} to \ref{ass4} hold, and that the interpolatory operator $P^N$ satisfies \eqref{thmreq}.
		\begin{enumerate}[label=(\roman*)]
			\item If $y^*$ has a nonzero index, then there exists an $N_0$ such that for $N \geq N_0$, $P^N \circ \mathcal{G} \circ \mathcal{T}$ has a fixed point $z^N \in V^N$ satisfying
			$$
			\Vert z^*-z^N \Vert \rightarrow 0 \qquad \mathrm{as} \quad N \rightarrow \infty.
			$$
			\label{thm6_1}
			\item Suppose in addition that \ref{ass5} holds, and that 1 is not an eigenvalue of the linear operator $D \left( \mathcal{G} \circ \mathcal{T} \right) (z^*)$.
			Then there exists a neighborhood of $z^*$ and an $N_1$ such that for $N \geq N_1$ a fixed point $z^N$ of $P^N \circ \mathcal{G} \circ \mathcal{T}$ is unique in that neighborhood, and
			$$
			c_2 \Vert z^*-P^Nz^* \Vert \leq  \Vert z^*-z^N \Vert \leq     c_3 \Vert z^*-P^Nz^* \Vert ,
			$$
			where $c_2,c_3>0 $ are independent of $N$.
			\label{thm6_2}
		\end{enumerate}
		\label{thmz}
	\end{theorem}
	An immediate corollary of Theorem \ref{thmz} establishes the optimal convergence of $z^N$, in the sense that it converges to $z^*$ with the same asymptotic order as the best approximation of $z^*$ in $V^N$.
	\begin{corollary}[Kumar \& Sloan \cite{kumar87}, Corollary 1]
		Suppose conditions \ref{ass1} to \ref{ass5} hold, and that 1 is not an eigenvalue of the linear operator $D \left( \mathcal{G} \circ \mathcal{T} \right) (z^*)$.
		Then there exists a constant $c_4>0$ such that
		$$
		\Vert z^* - z^N \Vert \leq c_4 \inf_{\phi \in V^N} \Vert z^*-\phi \Vert.
		$$
		\label{corollary0}
	\end{corollary}
	\begin{proof}
		The proof remains the same as the one given in \cite{kumar87}. Take $\phi \in V^N$, then
		$$
		\Vert z^*-P^Nz^* \Vert \overset{(\dagger)}{=} \Vert \left( \mathrm{id} - P^N \right) \left(z^*-\phi \right)  \Vert
		\leq \left( 1 + \Vert P^N \Vert \right) \Vert z^*-\phi \Vert ,
		$$
		where ($\dagger$) used that $P^N = \mathrm{id}$ on $V^N$.
		The result now follows from Theorem \ref{thmz}\ref{thm6_2} and the uniform boundedness of $P^N$.
	\end{proof}
	
	
	
	\section{Proof of Theorem \ref{thmy}}
	\label{appendix3}
	With the operators $\mathcal{G}$, $\mathcal{T}$ and $A$ defined in eqs.~\eqref{G}, \eqref{T} and \eqref{A}, we use the results stated in Appendix~\ref{appendix2} to prove Theorem~\ref{thmy}.  
	We first need to check that the assumptions \ref{ass1}-\ref{ass5} are satisfied by the integral equations \eqref{inteq} and eq.~\eqref{reforminteq}. Indeed, conditions \ref{ass1} and \ref{ass2} are automatically satisfied by the definition~\eqref{greenfunct} of the Green's function $L_m$. 
	
	The function~$L_m$ seems discontinuous due to the presence of step function $h(t)$, but since $h(t)$ is multiplied by a continuous function that takes the value 0 at $t=0$, $L_m$ is indeed continuous. The linear solution $\eta_{lin}$ is also continuous and $T$-periodic by Lemma~\ref{linearresponse} and the continuity of forcing $f(t)$. Hence, assumption \ref{ass3} is also satisfied.
	Furthermore, since the nonlinearity $S$ is of class $C^1$ in the hypothesis of Theorem~\ref{thmy}, assumptions~\ref{ass4} and \ref{ass5} also hold. 
	
	Finally, we note that assumption~\eqref{thmreq} is already satisfied for the piecewise polynomial interpolation operator  $P_N$~\eqref{interpop}~\cite{Vainikko1981}. This allows us to use Theorem~\ref{thmz}, which gives \eqref{thmy_firstassertion}.
    It follows from Lemma \ref{geomisollemma} that
	$$
	\eta^* = \mathcal{T} (\zeta^*)=f+A\zeta^*,
	$$
	thus
	$$
	\Vert \eta^*-\eta^N \Vert = \Vert A(\zeta^*-\zeta^N) \Vert \leq \Vert A \Vert 
	\left\Vert \zeta^*-\zeta^N \right\Vert.
	$$
	Now, since assumption \ref{ass1} means that
	$
	\Vert A \Vert  < \infty,
	$
	it follows that
	$$
	\Vert \eta^*-\eta^N \Vert \rightarrow 0 \qquad \mathrm{as} \quad N \rightarrow \infty,
	$$
	which proves \ref{thm3_1}.
	
	To show \ref{thm3_2}, we use Corollary \ref{corollary0} with the same argument as in part \ref{thm3_1} of the proof:
	$$
	\Vert \eta^*-\eta^N \Vert \leq \Vert A \Vert 
	\left\Vert \zeta^*-\zeta^N \right\Vert \leq c_4 \Vert A \Vert  
	\inf_{\phi \in V^N} \Vert \zeta^*-\phi \Vert = c \inf_{\phi \in V^N} \Vert \zeta^*-\phi \Vert.
	$$
	This concludes the proof of Theorem \ref{thmy}.
	
	\section{Proof of Theorem~\ref{picardthm}}
	\label{appendix4}
	Once we show that the conditions for the Banach fixed-point theorem are satisfied, the statement of Theorem~\ref{picardthm} follows directly.
	\renewcommand{\labelenumi}{\arabic{enumi}.}
	\begin{enumerate}
		\item $\left(C_{\zeta_0,\delta}^N, \Vert \, \cdot \, \Vert_{\infty}\right)$ is a closed subspace of the Banach space $C\left([0,T],\mathbb{R}^m \right)$, and is therefore a complete metric space.
		\item To show that $P^N \circ \mathcal{G} \circ \mathcal{T}: C_{\zeta_0,\delta}^N \rightarrow C_{\zeta_0,\delta}^N$ is satisfied, we need to prove that $P^N \circ \mathcal{G} \circ \mathcal{T} (\zeta^N)(t) = P^N \circ \mathcal{G} \circ \mathcal{T} (\zeta^N)(t+T)$ and $\Vert P^N \circ \mathcal{G} \circ \mathcal{T} (\zeta^N) - \zeta_0^N \Vert_{\infty} \leq \delta$.
		The first part is proven by a change of variables as follows
		\begin{align*}
			P^N \circ \mathcal{G} \circ \mathcal{T} (\zeta^N)(t) &= P^N \circ \mathcal{G} \left(
			\int_0^T L_m(t-s,T)  \left[ U_m^T f(s) + \zeta^N(s) \right] \mathrm{d}s
			\right)\\
			&\overset{(\dagger)}{=} P^N \circ \mathcal{G} \left(
			\int_0^T L_m(t+T-(s+T),T)  \left[ U_m^T f(s+T) + \zeta^N(s+T) \right] \mathrm{d}s
			\right)\\
			&= P^N \circ \mathcal{G} \left(
			\int_T^{2T} L_m(t+T-u,T)  \left[ U_m^T f(u) + \zeta^N(u) \right] \mathrm{d}u
			\right)\\
			&= P^N \circ \mathcal{G} \circ \mathcal{T} (\zeta^N)(t+T),
		\end{align*}
		where the equality ($\dagger$) follows from the periodicity of $f$ and $\zeta^N$.
		For the second part of the proof, we invoke assumption \ref{picardass1}:
		\begin{align*}
			\Vert P^N \circ \mathcal{G} \circ \mathcal{T} (\zeta^N) - \zeta_0^N \Vert_{\infty} &= \Vert P^N \circ \mathcal{G} \circ \mathcal{T} (\zeta^N) - P^N \circ \mathcal{G} \circ \mathcal{T} (\zeta_0^N) + P^N \circ \mathcal{G} \circ \mathcal{T} (\zeta_0^N) - \zeta_0^N \Vert_{\infty}\\
			&\leq \Vert P^N \circ \mathcal{G} \circ \mathcal{T} (\zeta^N) - P^N \circ \mathcal{G} \circ \mathcal{T} (\zeta_0^N) \Vert_{\infty} + \Vert P^N \circ \mathcal{G} \circ \mathcal{T} (\zeta_0^N) - \zeta_0^N \Vert_{\infty}\\
			&\leq \Vert P^N \Vert \Vert U_m^T \Vert L_S \Vert U_m \Vert \Vert \mathcal{T}(\zeta^N) - \mathcal{T}(\zeta_0^N) \Vert_{\infty} + \Vert \mathcal{E} (\zeta_0^N) \Vert_{\infty}\\
			&\leq \Vert P^N \Vert \Vert U_m^T \Vert L_S \Vert U_m \Vert \Vert A \Vert \delta + \Vert \mathcal{E} (\zeta_0^N) \Vert_{\infty} \leq \delta.
		\end{align*}
		\item $ P^N \circ \mathcal{G} \circ \mathcal{T}: C_{\zeta_0,\delta}^N \rightarrow C_{\zeta_0,\delta}^N$ is a contraction mapping, since for any $\zeta_1^N, \zeta_2^N \in C_{\zeta_0,\delta}^N$
		$$
		\Vert P^N \circ \mathcal{G} \circ \mathcal{T} (\zeta_1) - P^N \circ \mathcal{G} \circ \mathcal{T} (\zeta_2) \Vert_{\infty}
		\leq \Vert P^N \Vert \Vert U_m^T \Vert L_S \Vert U_m \Vert \Vert A \Vert \Vert \zeta_1-\zeta_2 \Vert_{\infty},
		$$
		where $ \Vert P^N \Vert \Vert U_m^T \Vert L_S \Vert U_m \Vert \Vert A \Vert < 1$ by assumption \ref{picardass2}.
	\end{enumerate}
	We can now apply the Banach fixed-point theorem to conclude the statement of Theorem~\ref{picardthm}.
	
	\clearpage
	\printbibliography

@article{Farhat2014,
	author = {Farhat, Charbel and Avery, Philip and Chapman, Todd and Cortial, Julien},
	doi = {10.1002/nme.4668},
	issn = {00295981},
	journal = {International Journal for Numerical Methods in Engineering},
	number = {9},
	pages = {625--662},
	publisher = {John Wiley {\&} Sons, Ltd},
	title = {{Dimensional reduction of nonlinear finite element dynamic models with finite rotations and energy-based mesh sampling and weighting for computational efficiency}},
	url = {http://doi.wiley.com/10.1002/nme.4668},
	volume = {98},
	year = {2014}
}

@article{Vainikko1981,
	author = {Vainikko, G. and Uba, P.},
	doi = {10.1017/s0334270000002770},
	issn = {0334-2700},
	journal = {The Journal of the Australian Mathematical Society. Series B. Applied Mathematics},
	number = {4},
	pages = {431--438},
	publisher = {Cambridge University Press (CUP)},
	title = {{A piecewise polynomial approximation to the solution of an integral equation with weakly singular kernel}},
	url = {https://doi.org/10.1017/S0334270000002770},
	volume = {22},
	year = {1981}
}

@article{Buza2020,
	archivePrefix = {arXiv},
	arxivId = {2009.04232},
	author = {Buza, Gergely and Jain, Shobhit and Haller, George},
	eprint = {2009.04232},
	title = {{Using Spectral Submanifolds for Optimal Mode Selection in Model Reduction}},
	url = {http://arxiv.org/abs/2009.04232},
	year = {2020}
}

@software{SST,
	author       = {Jain, Shobhit and
	Buza, Gergely and Breunung, Thomas and
	Haller, George},
	title        = {SteadyStateTool},
	publisher    = {Zenodo},
	version      = {v1.1},
	doi          = {10.5281/zenodo.3992820},
	url          = {https://doi.org/10.5281/zenodo.3992820},
	year = {2020}
}

@book{Dankowicz2013a,
	author = {Dankowicz, Harry and Schilder, Frank},
	booktitle = {Recipes for Continuation},
	doi = {10.1137/1.9781611972573},
	publisher = {Society for Industrial and Applied Mathematics, Philadelphia},
	title = {{Recipes for Continuation}},
	year = {2013}
}

@book{Krack2019,
	address = {Cham},
	author = {Krack, Malte and Gross, Johann},
	doi = {10.1007/978-3-030-14023-6},
	isbn = {978-3-030-14022-9},
	publisher = {Springer International Publishing},
	series = {Mathematical Engineering},
	title = {{Harmonic Balance for Nonlinear Vibration Problems}},
	url = {http://link.springer.com/10.1007/978-3-030-14023-6},
	year = {2019}
}

@article{Renson2016,
	author = {Renson, L. and Kerschen, G. and Cochelin, B.},
	doi = {10.1016/J.JSV.2015.09.033},
	issn = {0022-460X},
	journal = {Journal of Sound and Vibration},
	pages = {177--206},
	publisher = {Academic Press},
	title = {{Numerical computation of nonlinear normal modes in mechanical engineering}},
	url = {https://www.sciencedirect.com/science/article/pii/S0022460X15007543?via{\%}3Dihub},
	volume = {364},
	year = {2016}
}

@software{Auto,
	author = {E. J. Doedel and B. E. Oldeman and A. R. Champneys and F. Dercole and T. F. Fairgrieve and Y. Kuznetsov and R. C. Paffenroth and B. Sandstede and X. J. Wang and C. H. Zhang},
	title = { AUTO-07p: Continuation and bifurcation software for ordinary differential equations},
	url = {http://indy.cs.concordia.ca/auto/},
	version = {0.8},
	date = {2019-11-30},
}

@article{kumar87,
	ISSN = {00255718, 10886842},
	URL = {http://www.jstor.org/stable/2007829},
	author = {Kumar, Sunil and Sloan, Ian H.},
	journal = {Mathematics of Computation},
	volume = {48},
	number = {178},
	pages = {585--593},
	publisher = {American Mathematical Society},
	title = {A new collocation-type method for Hammerstein integral equations},
	doi = {10.2307/2007829},
	year = {1987}
}

@book{kantorovich1982functional,
	title={Functional Analysis},
	author={Kantorovich, L.V. and Akilov, G.P.},
	isbn={9780080264868},
	year={1982},
	publisher={Pergamon Press}
}

@book{krasnosel1984geometrical,
	title={Geometrical Methods of Nonlinear Analysis},
	author={Krasnoselski{\u\i}, M.A. and Zabre{\u\i}ko, P.P.},
	isbn={9783540129455},
	lccn={83016986},
	series={Grundlehren der mathematischen Wissenschaften},
	year={1984},
	publisher={Springer-Verlag Berlin Heidelberg}
}

@article{geomiso,
	ISSN = {00361429},
	doi = {10.1137/0718056},
	author = {Herbert B. Keller},
	journal = {SIAM Journal on Numerical Analysis},
	number = {5},
	pages = {822--838},
	publisher = {Society for Industrial and Applied Mathematics},
	title = {Geometrically isolated nonisolated solutions and their approximation},
	volume = {18},
	year = {1981}
}

@book{krasnosel1972approximate,
	title={Approximate Solution of Operator Equations},
	author = {Krasnosel'skii, M. A. and Vainikko, G. M. and Zabreiko, P. P. and Rutitskii, Ya. B. and Stetsenko, V. Ya.},
	isbn={9789401027175},
	series={Wolters-Noordhoff series on pure and applied mathematics},
	doi = {10.1007/978-94-010-2715-1},
	year={1972},
	publisher={Wolters-Noordhoff Publishing},
	address = {Groningen}
}

@book{geradin2015mechanical,
	title={Mechanical Vibrations: Theory and Application to Structural Dynamics},
	author={Geradin, M. and Rixen, D.J.},
	isbn={9781118900208},
	lccn={2014014588},
	year={2015},
	publisher={Wiley},
	address = {Chichester}
}

@Article{Jain2019,
	author="Jain, Shobhit
	and Breunung, Thomas
	and Haller, George",
	title="Fast computation of steady-state response for high-degree-of-freedom nonlinear systems",
	journal="Nonlinear Dynamics",
	year="2019",
	month="Jul",
	day="01",
	volume="97",
	number="1",
	pages="313--341",
	issn="1573-269X",
	doi="10.1007/s11071-019-04971-1",
	url="https://doi.org/10.1007/s11071-019-04971-1"
}

@book{buhler2018functional,
	title={Functional Analysis},
	author={B{\"u}hler, T. and Salamon, D.A.},
	isbn={9781470441906},
	lccn={2017057159},
	series={Graduate Studies in Mathematics},
	year={2018},
	publisher={American Mathematical Society},
	address = {Providence, Rhode Island}
}

@Article{Haller2016,
	author="Haller, George
	and Ponsioen, Sten",
	title="Nonlinear normal modes and spectral submanifolds: existence, uniqueness and use in model reduction",
	journal="Nonlinear Dynamics",
	year="2016",
	month="Nov",
	day="01",
	volume="86",
	number="3",
	pages="1493--1534",
	issn="1573-269X",
	doi="10.1007/s11071-016-2974-z",
	url="https://doi.org/10.1007/s11071-016-2974-z"
}

@article{JAIN2018195,
	title = "Exact nonlinear model reduction for a von Kármán beam: Slow-fast decomposition and spectral submanifolds",
	journal = "Journal of Sound and Vibration",
	volume = "423",
	pages = "195 - 211",
	year = "2018",
	issn = "0022-460X",
	doi = "https://doi.org/10.1016/j.jsv.2018.01.049",
	author = "Shobhit Jain and Paolo Tiso and George Haller"
}

@article{Ponsioen2019ExactMR,
	author = {Ponsioen, Sten and Jain, Shobhit and Haller, George},
	doi = {10.1016/j.jsv.2020.115640},
	issn = {10958568},
	journal = {Journal of Sound and Vibration},
	pages = {115640},
	publisher = {Academic Press},
	title = {{Model reduction to spectral submanifolds and forced-response calculation in high-dimensional mechanical systems}},
	volume = {488},
	year = {2020}
}

@book{nayfeh1,
	author = {Nayfeh, Ali Hasan},
	booktitle = {Perturbation Methods},
	doi = {10.1002/9783527617609},
	isbn = {9780471399179},
	publisher = {Wiley},
	address = {Weinheim},
	title = {{Perturbation Methods}},
	url = {https://onlinelibrary.wiley.com/doi/book/10.1002/9783527617609},
	year = {2000}
}

@article{Nayfeh1974NonlinearAO,
	author = {Nayfeh, Ali H. and Mook, Dean T. and Sridhar, Seshadri},
	doi = {10.1121/1.1914499},
	journal = {Journal of the Acoustical Society of America},
	number = {2},
	pages = {281--291},
	title = {{Nonlinear analysis of the forced response of structural elements}},
	url = {http://asa.scitation.org/doi/10.1121/1.1914499},
	volume = {55},
	year = {1974}
}

@article{10.1115/1.4040021,
	author = {Jain, Shobhit and Tiso, Paolo},
	title = "{Simulation-free hyper-reduction for geometrically nonlinear structural dynamics: a quadratic manifold lifting approach}",
	journal = {Journal of Computational and Nonlinear Dynamics},
	volume = {13},
	number = {7},
	year = {2018},
	month = {05},
	issn = {1555-1415},
	doi = {10.1115/1.4040021},
	url = {https://doi.org/10.1115/1.4040021},
	note = {071003},
}

@phdthesis{shellcode,
	author = {Tiso, Paolo},
	year = {2006},
	title = {Finite element based reduction methods for static and dynamic analysis of thin-walled structures},
	school = {Delft University of Technology}
}
\end{document}